\documentclass[fleqn,usenatbib]{mnras}

\usepackage{newtxtext,newtxmath}

\usepackage[T1]{fontenc}

\DeclareRobustCommand{\VAN}[3]{#2}
\let\VANthebibliography\thebibliography
\def\thebibliography{\DeclareRobustCommand{\VAN}[3]{##3}\VANthebibliography}

\usepackage{graphicx}	
\usepackage{amsmath}	

\title[The Obscured $z=7$ SFRD]{The ALMA REBELS Survey: The Dust-obscured Cosmic Star Formation Rate Density at Redshift 7}

\author[H.S.B. Algera et al.]{
Hiddo S. B. Algera,$^{1,2}$\thanks{E-mail: algera@hiroshima-u.ac.jp}
Hanae Inami,$^{1}$
Pascal A. Oesch,$^{3,4}$
Laura Sommovigo,$^{5}$
Rychard J. Bouwens,$^{6}$ 
\newauthor
Michael W. Topping,$^{7}$
Sander Schouws,$^{6}$
Mauro Stefanon,$^{6,8,9}$
Daniel P. Stark,$^{7}$
Manuel Aravena,$^{10}$
\newauthor
Laia Barrufet,$^{3}$
Elisabete da Cunha,$^{11,12}$
Pratika Dayal,$^{13}$
Ryan Endsley,$^{7}$
Andrea Ferrara,$^{5}$
\newauthor
Yoshinobu Fudamoto,$^{14,2}$
Valentino Gonzalez,$^{15}$
Luca Graziani,$^{16,17}$
Jacqueline A. Hodge,$^{6}$
\newauthor
Alexander P. S. Hygate,$^{6}$
Ilse de Looze,$^{18,19}$
Themiya Nanayakkara,$^{20}$
Raffaella Schneider,$^{16,17}$ and
\newauthor
Paul P. van der Werf$^{6}$
\\
$^{1}$Hiroshima Astrophysical Science Center, Hiroshima University, 1-3-1 Kagamiyama, Higashi-Hiroshima, Hiroshima 739-8526, Japan\\
$^{2}$National Astronomical Observatory of Japan, 2-21-1, Osawa, Mitaka, Tokyo, Japan\\
$^{3}$Observatoire de Gen\'{e}ve, 1290 Versoix, Switzerland \\
$^{4}$Cosmic Dawn Center (DAWN), Niels Bohr Institute, University of Copenhagen, Jagtvej 128, København N, DK-2200, Denmark \\
$^{5}$Scuola Normale Superiore, Piazza dei Cavalieri 7, I-56126 Pisa, Italy \\
$^{6}$Leiden Observatory, Leiden University, NL-2300 RA Leiden, Netherlands \\
$^{7}$Steward Observatory, University of Arizona, 933 N Cherry Ave, Tucson, AZ 85721, United States \\
$^{8}$Departament d'Astronomia i Astrof\`isica, Universitat de Val\`encia, C. Dr. Moliner 50, E-46100 Burjassot, Val\`encia,  Spain\\
$^{9}$Unidad Asociada CSIC "Grupo de Astrof\'isica Extragal\'actica y Cosmolog\'ia" (Instituto de F\'isica de Cantabria - Universitat de Val\`encia)\\
$^{10}$Nucleo de Astronomia, Facultad de Ingenieria y Ciencias, Universidad Diego Portales, Av. Ejercito 441, Santiago, Chile \\
$^{11}$International Centre for Radio Astronomy Research, University of Western Australia, 35 Stirling Hwy, Crawley,26WA 6009, Australia \\
$^{12}$ARC Centre of Excellence for All Sky Astrophysics in 3 Dimensions (ASTRO 3D), Australia \\
$^{13}$Kapteyn Astronomical Institute, University of Groningen, 9700 AV Groningen, The Netherlands \\
$^{14}$Waseda Research Institute for Science and Engineering, Faculty of Science and Engineering, Waseda University, 3-4-1 Okubo, Shinjuku, Tokyo 169-8555, Japan \\
$^{15}$Departamento de Astronomia, Universidad de Chile, Camino del Observatorio 1515, Las Condes, Santiago 7591245, Chile\\
$^{16}$Dipartimento di Fisica, Sapienza, Universita di Roma, Piazzale Aldo Moro 5, I-00185 Roma, Italy \\
$^{17}$INAF/Osservatorio Astronomico di Roma, via Frascati 33, I-00078 Monte Porzio Catone, Roma, Italy \\
$^{18}$Sterrenkundig Observatorium, Ghent University, Krijgslaan 281 - S9, 9000 Gent, Belgium \\
$^{19}$Dept. of Physics \& Astronomy, University College London, Gower Street, London WC1E 6BT, United Kingdom \\
$^{20}$Centre for Astrophysics \& Supercomputing, Swinburne University of Technology, PO Box 218, Hawthorn, VIC 3112, Australia \\
}

\date{Accepted XXX. Received YYY; in original form ZZZ}

\pubyear{2022}

\begin{document}
\label{firstpage}
\pagerange{\pageref{firstpage}--\pageref{lastpage}}
\maketitle


\begin{abstract}
Cosmic dust is an essential component shaping both the evolution of galaxies and their observational signatures. How quickly dust builds up in the early Universe remains an open question that requires deep observations at (sub-)millimeter wavelengths to resolve. Here we use Atacama Large Millimeter Array observations of 45 galaxies from the Reionization Era Bright Emission Line Survey (REBELS) and its pilot programs, designed to target [CII] and dust emission in UV-selected galaxies at $z\sim7$, to investigate the dust content of high-redshift galaxies through a stacking analysis. We find that the typical fraction of obscured star formation $f_\mathrm{obs} = \mathrm{SFR}_\mathrm{IR} / \mathrm{SFR}_\mathrm{UV + IR}$ depends on stellar mass, similar to what is observed at lower redshift, and ranges from $f_\mathrm{obs} \approx 0.3 - 0.6$ for galaxies with $\log_{10}\left(M_\star / M_\odot\right) = 9.4 - 10.4$. We further adopt the $z\sim7$ stellar mass function from the literature to extract the obscured cosmic star formation rate density (SFRD) from the REBELS survey. Our results suggest only a modest decrease in the SFRD between $3\lesssim z \lesssim 7$, with dust-obscured star formation still contributing $\sim30\%$ at $z\sim7$. While we extensively discuss potential caveats, our analysis highlights the continued importance of dust-obscured star formation even well into the epoch of reionization.
\end{abstract}



\section{Introduction}
\label{sec:introduction}

The cosmic star formation rate density (SFRD) is one of the most fundamental quantities describing the evolution of the galaxy population across time. Within the last decade it has been established that the SFRD reaches its maximum around a redshift of $z\sim1-3$, while since having been on the decline towards the present day (\citealt{madau2014} and references therein). However, the nature of the SFRD in the early Universe,  especially beyond $z\gtrsim5$, remains largely unconstrained. Through wide surveys combining ground- and space-based facilities, including in particular the \emph{Spitzer} Space Telescope, ever-increasing samples of high-redshift ($z\gtrsim6$) galaxies are being discovered (e.g., \citealt{mclure2013,bouwens2015,finkelstein2015,oesch2018,stefanon2019}). However, such studies, for the most part, remain limited to probing the rest-frame ultraviolet (UV) and optical emission emanating from distant galaxies, and in turn only provide constraints on the light directly emitted by stars. This is even true about some new samples now being compiled based on new James Webb Space Telescope (JWST) data (e.g., \citealt{adams2022,harikane2022,naidu2022}). 

Meanwhile, at lower redshift it has long been established that cosmic dust plays a significant role in shaping the observable properties of galaxies (e.g., \citealt{draine1989,draine2003,calzetti2000}). Dust, itself being a by-product of star formation, attenuates starlight at UV and optical wavelengths, and reprocesses it into infrared emission. Both locally and at intermediate redshifts ($1 \lesssim z \lesssim 4$), the most actively star-forming galaxies are known to be enshrouded in large amounts of dust (e.g., \citealt{kennicutt1998,casey2014,dudzeviciute2020}). In addition, the SFRD is known to be dominated by dust-obscured star formation at least out to $z\lesssim4$ \citep{magnelli2011,magnelli2013,gruppioni2013,gruppioni2020,bouwens2016,dunlop2017,zavala2021}. Therefore, in order to assess the dust content and star formation properties of more distant galaxies,  observations at (far-)infrared wavelengths are essential.

Throughout the last decade, the Atacama Large Millimeter Array (ALMA) has revolutionized the study of both the dust and interstellar medium properties (ISM) of distant galaxies (see \citealt{hodge2020} for a review). In recent years, a handful of $z\gtrsim7$ galaxies have been identified in dust emission (e.g., \citealt{watson2015,laporte2017,laporte2019,bowler2018,tamura2019,bakx2020,bakx2021,schouws2021,sugahara2021,endsley2022,witstok2022}), thereby implying that significant dust reservoirs are already in place in the early Universe. 

However, given that direct observations of dust at high redshift remain scarce, the importance of dust-obscured star formation in the typical galaxy population within the epoch of reionization ($z\gtrsim 7$; \citealt{planck2020}) remains unclear. While UV-based studies of cosmic star formation generally measure a steep decline in the SFRD beyond $z\gtrsim3$ \citep{madau2014,bouwens2015,finkelstein2015}, such results appear in tension with recent measurements of dust-obscured star formation at $z\sim3-5$ (e.g., \citealt{novak2017,bouwens2020,dudzeviciute2020,gruppioni2020,khusanova2021,vandervlugt2022}). This discrepancy may plausibly be the result of underestimated dust-corrections applied to UV-based studies, which are generally calibrated on low-redshift starburst galaxies (e.g., \citealt{meurer1999}). In addition, the existence of so-called optically dark galaxies (e.g., \citealt{simpson2014,wang2019,algera2020b,gruppioni2020,zhou2020,smail2021,talia2021,fudamoto2021,gomezguijarro2022}), which evade detection even in deep UV- and optical imaging, are impossible to account for in UV-based surveys of cosmic star formation without ancillary infrared observations. 

It is therefore clear that systematic surveys of dust in the early Universe are crucial for extending our knowledge of cosmic star formation into the epoch of reionization. Such a systematic approach can provide us with insight into the typical fraction of dust obscured star formation in galaxies at high redshift, which has currently been measured out to $z\sim5.5$ \citep{whitaker2017,fudamoto2020}. At high redshift, there is ongoing discussion about the contribution of different dust production mechanisms (i.e., that produced by Type II supernovae and AGB stars) versus the importance of grain growth in the interstellar medium  \citep{todini2001,mancini2015,popping2017,lesniewska2019,vijayan2019,dayal2022}. Hence, by exploring dust properties of galaxies in the early Universe, we can attempt to constrain the possible mechanisms and corresponding efficiencies of dust build-up at high redshift (e.g., \citealt{graziani2020,dayal2022,ferrara2022,sommovigo2022}).


In this paper, we investigate the history of obscured cosmic star formation using the ALMA Reionization Era Bright Emission Line Survey (REBELS; \citealt{bouwens2021}). REBELS is being executed as a Cycle-7 ALMA Large Program and represents the first systematic survey of the ISM and dust properties of $z\gtrsim6.5$ galaxies. In Section \ref{sec:data} we briefly introduce the REBELS ALMA observations. In Section \ref{sec:analysis} we introduce our stacking routine, as well as our methodology of determining the $z\sim7$ obscured cosmic SFRD with a targeted sample. In Section \ref{sec:results_fobs} we present the typical fraction of dust-obscured star formation in $z\sim7$ UV-selected galaxies, while in Section \ref{sec:results_sfrd} we show our direct constraints on the obscured SFRD. Finally, we summarize our results in Section \ref{sec:summary}. Throughout this work, we assume a standard $\Lambda$CDM cosmology, with $H_0=70\,\text{km\,s}^{-1}\text{\,Mpc}^{-1}$, $\Omega_m=0.30$ and $\Omega_\Lambda=0.70$ and adopt a \citet{chabrier2003} initial mass function.

\section{Data}
\label{sec:data}

\subsection{REBELS}
The REBELS survey is an ongoing ALMA Large Program aimed at identifying particularly luminous ISM reservoirs at $z\gtrsim6.5$ through the detection of the [CII]$158\,\mu$m or [OIII]$88\mu$m lines and dust emission in 40 UV-selected galaxies. For details on the sample selection and observational strategy, we refer the reader to \citet{bouwens2021}, while the calibration and data reduction are described in detail in Schouws et al. (2022, in prep.). In summary, the REBELS targets were selected to be UV-bright galaxies ($M_\mathrm{UV} < -21.3\,$mag) with robustly constrained photometric redshifts placing them within the range $6.5 \lesssim z \lesssim 9.5$. 

The galaxies were drawn from a variety of well-studied extragalactic fields, spanning a total area of $\sim7\,\mathrm{deg}^2$. The majority of the REBELS sources were selected across the well-studied COSMOS field \citep{scoville2007}, which includes photometric coverage in the \emph{BgVriz} bands from the Canda-France-Hawaii Telescope (CFHT), the \emph{ugrizy} bands from Subaru/Suprimecam and the YJHK$_s$ bands from the UltraVISTA survey \citep{mccracken2012}, as well as \emph{Spitzer}/IRAC coverage in Channels 1 and 2. The bulk of the remaining targets are drawn from the UKIDSS/UDS and VIDEO/XMM-LSS surveys \citep{lawrence2007}, and have similar available photometric coverage, as described in detail in \citet{bouwens2022}.

For each of the selected targets, the photometric redshift probability distribution was scanned for either the [CII] ($z \leq 8.2$; 36 sources) or [OIII] ($z > 8.2$; 4 sources) emission line using between two and six ALMA spectral scans. Based on the photometric redshift, either ALMA Bands 5, 6 or 7 were used to scan down to a line sensitivity of roughly $L_\mathrm{[CII]} \approx 2\times10^8\,L_\odot$ for a galaxy at $z=7$ ($5\sigma$). As a result of the variable number of scans used, the typical continuum sensitivity of the various ALMA pointings ranges from $\sigma = 10 - 25\,\mu\mathrm{Jy\,beam}^{-1}$. All observations were taken in one of two compact ALMA configurations (C43-1, C43-2), resulting in a typical angular resolution of $1\farcs2 - 1\farcs6$. In this work, we focus only on the 36/40 galaxies targeted in [CII] emission, for which the underlying rest-frame $158\,\mu$m continuum is observed. This sample does not include the two serendipitously detected galaxies described in \citet{fudamoto2021}, given their different selection criteria and the lack of a detection at rest-frame UV wavelengths.

During Cycle 7, $60.6$ out of the allocated $70$ hours have been executed, resulting in $23$ detections of [CII] emission above $5\sigma$ (Schouws et al. 2022, in prep.) as well as $14$ detections of the dust continuum at rest-frame $158\,\mu$m above a threshold of $3.3\sigma$ \citep{inami2022}.\footnote{Note that there are two additional continuum detections at $88\,\mu$m, increasing the total to 16 sources. However, in this work we only analyze the REBELS sample targeted at rest-frame $158\,\mu$m.}

In addition, we include $9$ sources from the REBELS pilot projects (\citealt{smit2018,schouws2021,schouws2022a}) in our analysis. Similar to REBELS, these sources were observed via ALMA spectral scans with the aim of detecting the [CII] line and underlying continuum emission. Among the 9 pilot sources, two are continuum-detected and five have [CII]-based spectroscopic redshifts. In total, our sample therefore consists of 45 galaxies, 16 of which are individually detected in dust continuum emission. For further details on the dust continuum properties of the REBELS sources, we refer to \citet{inami2022}.

Throughout this work, we will repeatedly compare our results to those obtained from the ALMA Large Program to Investigate C+ at Early Times (ALPINE; \citealt{lefevre2020,bethermin2020,faisst2020}). ALPINE is similar to REBELS in that it obtains ALMA observations of [CII] and dust emission for UV-selected galaxies, albeit for a lower redshift sample spanning the range of $4 \lesssim z \lesssim 6$ and with prior spectroscopic redshifts available. Given the similarities in sample selection, ALPINE forms the natural comparison sample to REBELS at lower redshift.

\subsection{Stellar Masses}
\label{sec:Mstar}
The total stellar mass ($M_\star$) is one of the principal parameters characterizing a galaxy, directly related to its integrated star formation history. However, measuring stellar masses at high redshift is difficult owing to the generally small number of photometric datapoints, scarce coverage of rest-frame optical and near-infrared emission, potentially significant uncertainties in photometric redshifts, and possible impact of emission lines (e.g., \citealt{stark2013,smit2015,topping2022}). Given the large fraction of [CII]-detections in REBELS, some of these uncertainties are partially mitigated. Nevertheless, determining stellar masses relies on a variety of assumptions, such as galaxies' star formation histories (SFHs) and the adopted dust attenuation curve, and therefore involves several inherent uncertainties.

Two separate works describe the determination of stellar masses for the REBELS sample, adopting two different methodologies. Stefanon et al. (2022, in prep.) adopt the spectral energy distribution (SED) fitting code {\sc{beagle}} \citep{chevallard2016} in combination with a constant SFH and an SMC dust law to determine the physical properties of the REBELS sample. However, the adoption of a constant SFH is known to result in very young ages for galaxies caught in the midst of a starburst (e.g., \citealt{leja2019}), as these outshine any older stellar populations that may already be present. To account for this possibility, \citet{topping2022} additionally present stellar masses derived using a non-parametric SFH for the REBELS sample using the {\sc{Prospector}} SED fitting code \citep{johnson2021}. Except for the different SED fitting routine and stellar population templates, their assumptions are identical to those used in Stefanon et al. (2022, in prep.). We briefly summarize these assumptions here, while we refer to \citet{topping2022} for the full details (see also \citealt{whitler2022}). In short, the SED fitting utilizes all available optical to mid-infrared photometry (Section \ref{sec:data} and \citealt{bouwens2022}), including any available narrow-band photometry. The Flexible Stellar Population Synthesis (FSPS) templates \citep{conroy2009,conroy2010} are adopted, as well as a Small Magellanic Cloud dust attenuation law. In total, eight time bins are used to construct the non-parametric star formation histories, with the oldest bin extending to $z=20$. The first two bins have fixed ages of $0 - 3$ and $3 - 10\,$Myr, and are particularly important to constrain potential strong rest-frame optical line emission that can boost the observed fluxes in the IRAC channels. Such nebular emission is included through the models from \citet{byler2017}. The remaining six bins are distributed logarithmically in time, and all bins are connected through the built-in ``continuity prior'' in {\sc{Prospector}} which downweights significant SFR variations between consecutive bins. From the {\sc{Prospector}} fits, \citet{topping2022} find that the non-parametric models prefer larger stellar masses, with an average offset of $0.43\,$dex, compared to those derived using a constant SFH. The typical offset is largest for young galaxies, in which case the difference can be up to $1.0\,$dex.

Given the greater flexibility offered by the non-parametric modelling, we adopt the \citet{topping2022} stellar masses for the REBELS galaxies throughout this work. Based on these masses, the majority of the REBELS sample falls onto the \citet{schreiber2015} star formation main sequence extrapolated to $z=7$ (Figure \ref{fig:mainsequence}; Section \ref{sec:starburst}). Given the inherent uncertainties associated with stellar mass determinations, we propagate the errors on the individual galaxy masses throughout our analysis using an MC-based approach, to ensure these are encapsulated in the quoted uncertainties on the quantities of interest, such as the obscured SFRD. Further details are provided in the following Section \ref{sec:analysis}.

\section{Analysis}
\label{sec:analysis}

\subsection{Stacking}
\label{sec:stacking}
While a significant fraction ($\sim40\%$) of REBELS galaxies are individually detected in the dust continuum \citep{inami2022}, in this work we also include the continuum-undetected sources through a stacking analysis in order to provide a complete census of dust in UV-selected galaxies at $z\sim7$. 

We utilize mean stacking in the image plane, at the native resolution of the REBELS images. We adopt the observed UV positions of the REBELS targets which are not individually detected in the ALMA maps, while we use the central position at rest-frame $158\,\mu$m for the dust detected sources. Simultaneously, we stack empty regions in the ALMA pointings prior to correcting for the primary beam attenuation to obtain a census of the (flat) noise properties of the stacked image. In addition, we verify the background level is negligible in both the individual ALMA pointings and in the stacks themselves. In order to identify emission in the stacks, we follow \citet{algera2022} by running {\sc{PyBDSF}} \citep{mohanrafferty2015} on the stacked images with peak and island thresholds of $3\sigma$ and $2\sigma$, respectively, where $\sigma$ is the RMS noise in the stack. In this work, we adopt the stacked peak flux densities, as integrated flux measurements may be biased upwards for low signal-to-noise stacks \citep{algera2020a,leslie2020}. Given that dust continua in high-redshift galaxies appear to be compact and are therefore not resolved by the large ($\sim1\farcs3$) REBELS beam size \citep{inami2022,fudamoto2022}, we do not expect this to substantially affect our flux densities. In addition, the beam sizes are significantly larger than the typical observed offset between the UV and dust continuum for the REBELS dust-detected sample ($\sim0\farcs3$; \citealt{inami2022}). In the case of a non-detection in the stacks, we follow the approach by \citet{bethermin2020} and adopt as a conservative upper limit $\max\{3\sigma,3\sigma + S_\mathrm{pxl}\}$. More specifically, we add the flux density in the central pixel of the stack $S_\mathrm{pxl}$ to the $3\sigma$ upper limit, provided that $S_\mathrm{pxl} > 0$. We further note that two bright, serendipitous galaxies $\sim5''$ from the image centers of REBELS-05 and REBELS-P4 were subtracted via {\sc{PyBDSF}} in order to avoid contaminating the background level in the stacks.
\\

\begin{figure}
    \centering
    \includegraphics[width=0.5\textwidth]{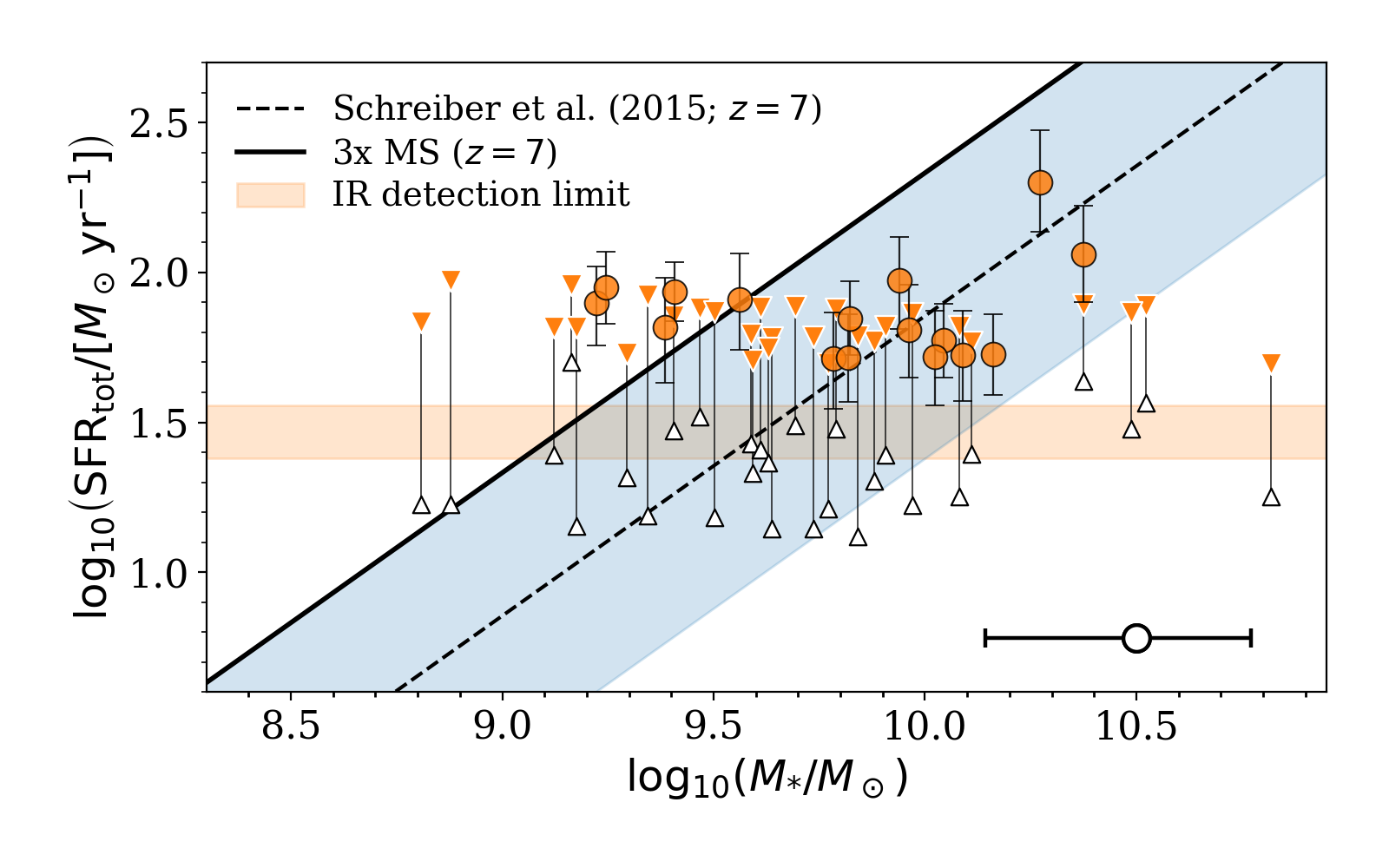}
    \caption{Position of the REBELS sources on the \citet{schreiber2015} main sequence, extrapolated to $z=7$. Orange circles indicate the dust-detected REBELS sources, while non-detections are bounded by their UV SFRs (white upward triangles) and the combined UV SFRs and IR upper limits (orange downward triangles). The median uncertainty on the galaxy stellar masses is indicated by the errorbar in the lower right corner. The blue shaded area shows the range of $\log_{10}\left(\mathrm{SFR}_\mathrm{MS}\right) \pm \log_{10}\left(3\right)$ at fixed stellar mass. Galaxies with total ($=$ UV + IR) SFRs exceeding $3\times\mathrm{SFR}_\mathrm{MS}$, or with UV SFRs exceeding this value in the case of a dust non-detection, are identified as starbursts (Section \ref{sec:starburst}). The orange band indicates the typical REBELS IR detection limit of $L_\mathrm{IR} \approx 2 - 3\times10^{11}\,L_\odot$ \citep{inami2022}. At low stellar masses, only galaxies above the (extrapolated) main sequence are likely to be individually IR-detected.}
    \label{fig:mainsequence}
\end{figure}

One of the aims of this work is to determine the typical fraction of obscured star formation in UV-selected galaxies at $z\sim7$. At $z\lesssim2.5$, this fraction is known to be highly mass-dependent (e.g., \citealt{pannella2009,reddy2010,whitaker2017}), which we investigate with REBELS by dividing our sample into discrete stellar mass bins. However, this is complicated by the stellar mass uncertainties (typical $1\sigma$ error of $\sim0.3\,$dex), which constitute a significant fraction of the mass range spanned by REBELS ($\log_{10}\left(M_\star / M_\odot\right) = 8.8 - 10.8$; interquartile range of $\log_{10}\left(M_\star / M_\odot\right) = 9.4-10.0$; see also Figure \ref{fig:mainsequence}). Therefore, we additionally incorporate these uncertainties into our analysis, instead of adopting a fixed stellar mass for our sources. We approximate the posterior probability distributions on $M_\star$ as a two-sided Gaussian with standard deviations equal to the lower and upper $1\sigma$ uncertainty on the median $M_\star$ derived from SED fitting by \citet{topping2022}, and randomly sample $N_\mathrm{MC} = 100$ times from this distribution for each of the galaxies we are stacking. For each of the $N_\mathrm{MC}$ samplings, we bin the data into four distinct bins with an approximately equal number of galaxies ($8-10$), and adopt these bins for stacking. In turn, a galaxy may end up in different bins across the $N_\mathrm{MC}$ stacking runs. In addition, the median $\langle M_\star \rangle$ of each bin differs slightly between the stacking runs. We subsequently perform photometry on all the stacks in order to determine the average $158\,\mu\mathrm{m}$ flux density in each bin. For a more direct comparison with the literature, where stellar mass uncertainties are generally not taken into account when stacking, we also provide ``default'' stacks throughout this work. In these default stacks, all galaxies are taken to have a fixed $M_\star$ equal to the median of the posterior stellar mass probability distribution. In this case, we use a standard bootstrap analysis (following \citealt{algera2020a}) such that the quoted uncertainties reflect both the error on the stacked photometry and the intrinsic sample variance. We refer to these as the ``no-MC'' stacks throughout this work.

\subsection{Infrared Luminosities and SFRs}
\label{sec:analysis_Lir}
We correct the stacked flux densities for the cosmic microwave background following \citet{dacunha2013} (see also \citealt{inami2022}), thereby adopting the median redshift of the stack. In order to convert the flux densities to infrared luminosities, we adopt the conversion from \citet{sommovigo2022} derived specifically for REBELS. Their models combine the [CII] and dust continuum emission of the $13$ REBELS targets for which both are available with the relation between [CII] luminosity and star formation rate (e.g., \citealt{delooze2014}) to constrain the shape of their infrared SEDs, and hence their infrared luminosities. As detailed in \citet{inami2022}, the median SED across the 13 sources analyzed by \citet{sommovigo2022} is adopted for all REBELS targets, resulting in a fixed conversion factor between rest-frame $158\,\mu$m dust continuum flux density and total infrared luminosity. However, this conversion factor shows a minor dependence on stellar mass, as the models from \citet{sommovigo2022} limit the maximum possible dust mass to be 4\% of the galaxy stellar mass. In this work, we therefore adopt a slightly different conversion factor, adapted for the non-parametric stellar masses from \citet{topping2022}. The infrared luminosities are calculated as $L_\mathrm{IR} = 11.7_{-2.3}^{+4.1} \times \nu_0 L_{\nu_0}$, where $\nu_0 \approx 1900\,$GHz is the rest-frame frequency of the [CII] emission line and $ L_{\nu_0}$ the specific continuum luminosity at this frequency \citep{sommovigo2022}.\footnote{Due to the non-parametric masses generally exceeding their parametric counterparts, the \citet{sommovigo2022} models allow for more dust when the former masses are adopted. As a result, the average dust temperature of the REBELS sample decreases slightly (by $\Delta T_\mathrm{dust} \approx 3\,$K), well within the uncertainties of the median temperature found in \citet{sommovigo2022} of $\langle T_\mathrm{dust} \rangle = 47 \pm 6\,$K when the parametric masses are adopted.} The uncertainty on this conversion factor is propagated into the final uncertainties on all quantities derived from the infrared luminosity. In addition, when determining dust-obscured star formation rates from $L_\mathrm{IR}$, we adopt the conversion factor used throughout the REBELS papers given in \citet{bouwens2021} of $\mathrm{SFR}_\mathrm{IR} / (M_\odot\,\mathrm{yr}^{-1}) = 1.2 \times 10^{-10} \times \left(L_\mathrm{IR} / L_\odot\right)$. This value -- as well as the conversion for UV-based star formation rates quoted in Section \ref{sec:starburst} -- is obtained from \citet{madau2014} under the assumption of a constant star formation rate for a duration of 100\,Myr and a fixed metallicity of $Z=0.002$ (see also Figure 3 in \citealt{madau2014}). While the precise conversion factor between star formation rate and IR (or UV) luminosity depends on the assumed star formation history -- and is therefore both time- and model-dependent -- we here adopt a fixed value for simplicity (see also \citealt{topping2022,whitler2022}). Given that we utilize stacking, we average across galaxies of various ages, further motivating the assumption of a fixed conversion factor corresponding to the average age of the REBELS sample of $100\,$Myr \citep{topping2022}.

\begin{figure*}
    \centering
    \includegraphics[width=0.99\textwidth]{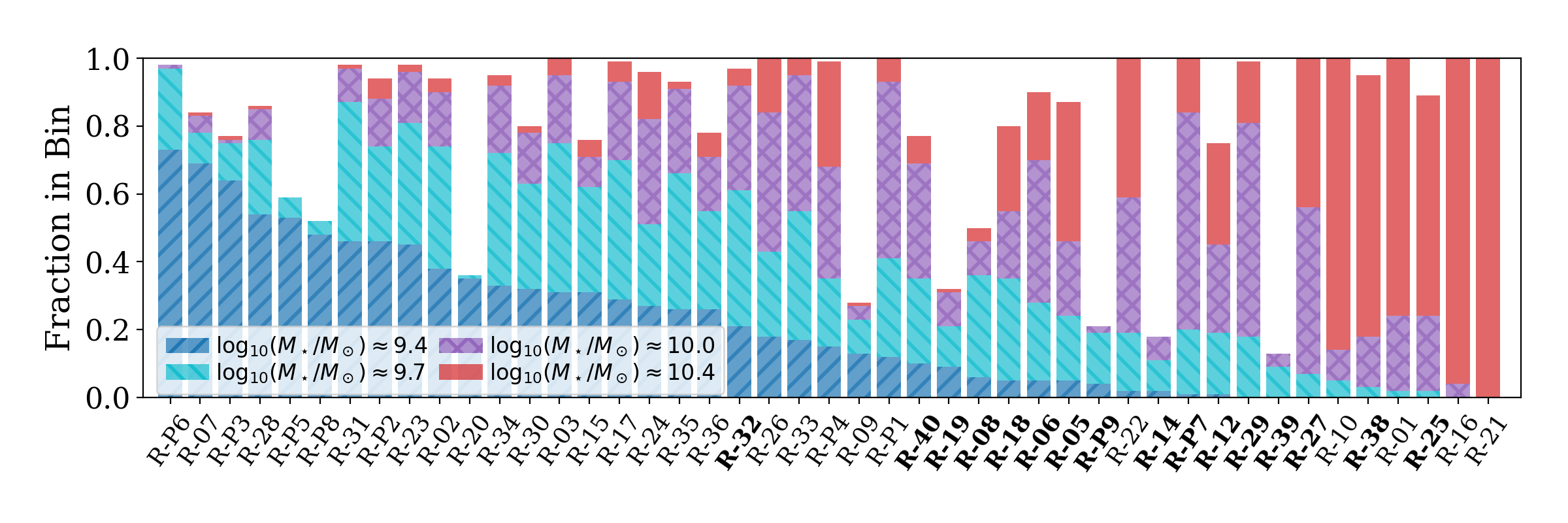}
    \caption{An illustration of the Monte Carlo binning process, showcasing the fraction of runs in which a galaxy lands in a given bin. Galaxies are ordered by decreasing incidence in the low mass bins, and dust-detected sources are highlighted in boldface. In cases where the bars do not sum to unity, a galaxy is identified as a starburst in the remaining fraction of MC runs (Section \ref{sec:starburst}). While some galaxies (e.g., R-14) are identified as starbursts in nearly all MC runs, several others (e.g., R-25) are flagged as starbursts only in a small fraction of runs when assigned a low stellar mass. Overall, roughly half of sources end up in any single bin more than half of the time, indicating that the typical stellar mass uncertainties of the REBELS targets will propagate into the eventual determination of the obscured SFRD in Section \ref{sec:results_sfrd}.}
    \label{fig:fraction_in_bin}
\end{figure*}

\subsection{Starburst Identification}
\label{sec:starburst}

In order to construct an unbiased sample of representative galaxies at $z\sim7$, we seek to identify and flag starburst galaxies in this work. The reason for this is twofold: in Section \ref{sec:results_fobs} we investigate how the fraction of obscured star formation depends on galaxy physical parameters such as stellar mass. To place this in the context of previous work, we compare to the $z\sim5.5$ ALPINE sample \citep{fudamoto2020}, which represents typical main-sequence galaxies, and the low-redshift sample from \citet{whitaker2017} ($z\lesssim2.5$). The latter, being stellar mass-complete by construction, also contains only a small number of starbursts (roughly $2\%$ by number; \citealt{rodighiero2011}). In addition, we seek to construct the $z\sim7$ obscured cosmic SFRD in Section \ref{sec:results_sfrd}, which necessitates first establishing a relation between infrared luminosity and stellar mass (Section \ref{sec:analysis_sfrd}). As determining the obscured SFRD requires extrapolating this relation to masses outside the range probed by REBELS, it is crucial the relation be representative for galaxies across a wide range of stellar masses, further motivating the exclusion of starbursts. \\

The REBELS targets were selected based on their UV luminosities, and as such were not explicitly required to reside on the main sequence. Given that a set number of ALMA spectral scans were used for each target, the continuum sensitivities of the REBELS pointings depend only weakly on redshift. As a result of this approximately constant sensitivity to dust-obscured star formation, REBELS is mostly sensitive to starbursts at low masses ($\log_{10} (M_\star / M_\odot) \lesssim 9.5$). At higher masses, galaxies on the (extrapolated to $z=7$) main sequence are also selected (Figure \ref{fig:mainsequence}).

We identify starbursts based on their position above the star-forming main sequence, as is typical in the literature (e.g., \citealt{rodighiero2011,caputi2017}). This classification is appealing from an observational perspective, as it is agnostic to any potentially different physical conditions in starburst and main-sequence galaxies, relating to, for example, merger state (e.g., \citealt{kartaltepe2012}) and gas fraction or star formation efficiency (e.g., \citealt{scoville2016,tacconi2018}).

We flag galaxies as starbursts if their combined UV + IR SFR places them above $3\times$ the \citet{schreiber2015} main sequence, extrapolated to $z=7$ (Figure \ref{fig:mainsequence}). Our sample is -- by construction -- rest-frame UV-detected, with its UV luminosities having been determined by Stefanon et al. (2022, in prep.), who additionally adopt the conversion factor to UV star formation rate of $\mathrm{SFR}_\mathrm{UV} / (M_\odot\,\mathrm{yr}^{-1})= 7.1\times10^{-29} \times \left(L_\mathrm{UV} / (\mathrm{erg/s/Hz})\right)$.\footnote{We note that the SFRs derived by two different SED fittings, Stefanon et al. (2022, in prep.) and \citet{topping2022} (averaged across 10\,Myr), are in excellent agreement. Indeed, \citet{topping2022} show that the choice of star formation history has minimal effect on the inferred unobscured SFR.} Galaxies without an IR detection are identified as starbursts when their UV SFRs alone place the galaxies $3\times$ above the main sequence, irrespective of their (unknown) IR properties. When performing our Monte Carlo stacking analysis, we flag starbursts based on their re-sampled stellar mass (Figure \ref{fig:fraction_in_bin}). In turn, a galaxy may be identified as a starburst only in a subset of the stacking runs. This analysis excludes, on average, $8.5 \pm 2.1$ out of the 45 REBELS sources from stacking. We verified that adopting the main sequence parametrization from \citet{speagle2014} results in a similar set of galaxies being identified as starbursts, and hence the precise functional form of the main sequence does not affect our results (see also Figure 8 in \citealt{topping2022}). However, we acknowledge the inherent uncertainties associated with extrapolating these main sequences to $z=7$, and therefore we discuss the effect of omitting the starburst exclusion in detail in Section \ref{sec:results_sfrd_starbursts}.

\subsection{The Obscured Cosmic Star Formation Rate Density}
\label{sec:analysis_sfrd}

In this work, we use REBELS to constrain the obscured cosmic star formation rate density at $z\sim7$. However, determining the obscured SFRD using the standard approach of constructing the infrared luminosity function is complicated by the targeted nature of the REBELS sample. Nevertheless, using the more robustly established UV luminosity function at $z=7$ as a proxy, \citet{barrufet2022} present constraints on the IR luminosity function from REBELS. In this work, however, we follow the procedure of the ALPINE survey -- which, like REBELS, is a pre-selected survey -- presented in \citet{khusanova2021}. The procedure is as follows: we adopt the recent determination of the galaxy stellar mass function (SMF; denoted as $\Phi(M_\star)$) at $z=7$ from \citet{stefanon2021}, thereby including the uncertainties on their derived SMF parameterization. For consistency with the \citet{chabrier2003} IMF adopted throughout this work, we shift their SMF, which assumes a \citet{salpeter1955} IMF, to lower stellar masses by $0.23\,$dex, following \citet{speagle2014}. We stack the REBELS sample in distinct mass bins in order to determine a relation between the stacked infrared luminosity $L_\mathrm{IR}$ and stellar mass $M_\star$. The infrared luminosity density $\rho_\mathrm{IR}$ is then given by

\begin{align}
    \rho_\mathrm{IR} = \int_{M_\star^\mathrm{lower}}^{M_\star^\mathrm{upper}} \Phi(L_\mathrm{IR}(M_\star)) L_\mathrm{IR}(M_\star) d M_\star \ ,
    \label{eq:rho_ir}
\end{align}

\noindent where the integral is calculated across a range of stellar masses between $[M_\star^\mathrm{lower}, M_\star^\mathrm{upper}]$. Suitable values for this range are discussed in detail in Section \ref{sec:sfrd_full}. The dust-obscured luminosity density $\rho_\mathrm{IR}$ is subsequently converted into a star formation rate density using the conversion factor from Section \ref{sec:analysis_Lir}. In this work, we follow the approach in \citet{khusanova2021}, but also extend their analysis by propagating the uncertainties on the stellar masses, which are generally not taken into account in the literature. In turn, we additionally shed light on how such uncertainties eventually propagate into the quantity of interest, the obscured cosmic SFRD.

\section{The Fraction of Obscured Star Formation}
\label{sec:results_fobs}

\subsection{Obscuration across the full REBELS sample}

\begin{figure}
    \centering
    \includegraphics[width=0.45\textwidth]{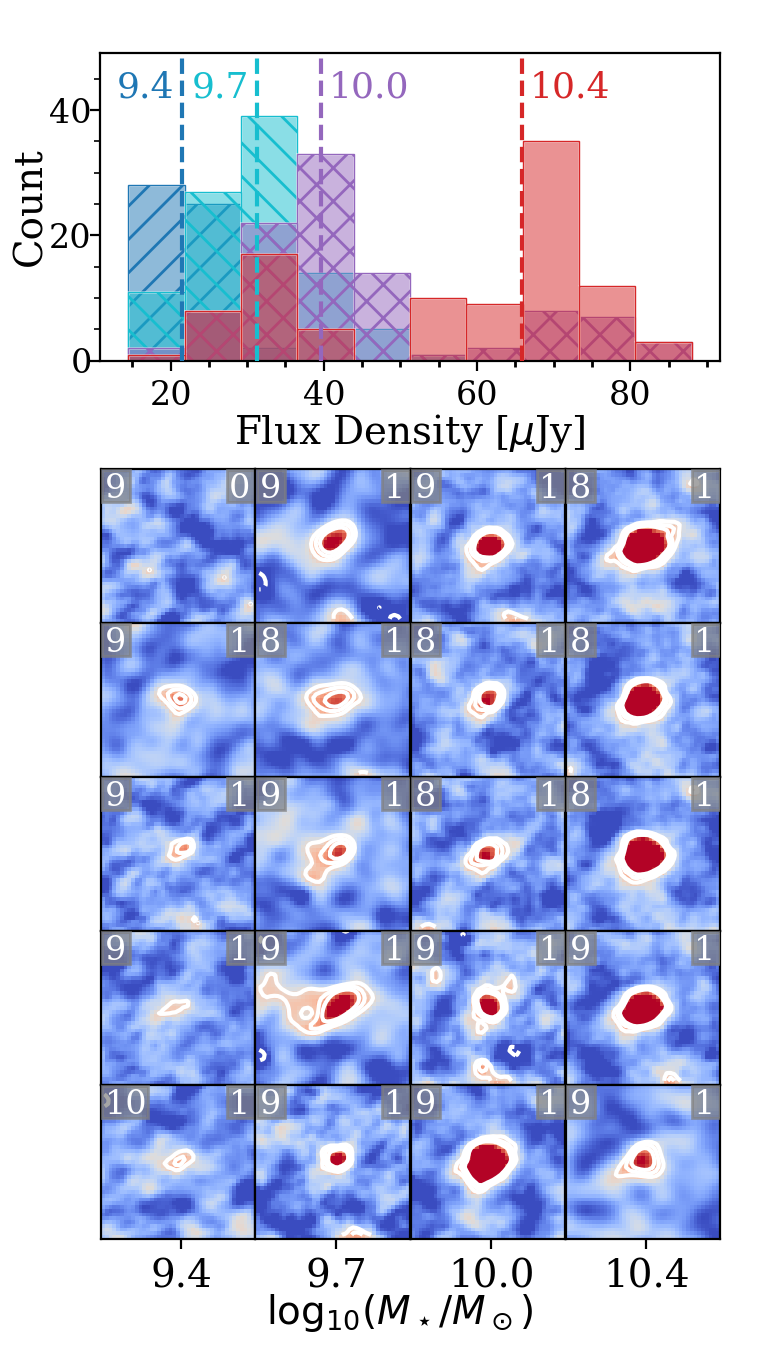}
    \caption{An illustration of the Monte Carlo stacking process. \textbf{Upper:} histogram of the rest-frame $158\,\mu$m flux densities across all $N_\mathrm{MC} = 100$ stacking runs, divided across the four stellar mass bins. Dashed lines correspond to the median flux density across the bins with stacked detections of dust continuum emission. The median mass of each bin is labeled. \textbf{Lower:} Randomly drawn subset of $N_\mathrm{MC} \times N_\mathrm{bins} = 5\times4$ Monte Carlo stacks (approximately $5''\times5''$) showing the rest-frame $158\,\mu$m continuum emission, ordered from low- (left) to high mass (right). The $x-$axis is labeled with the median stellar mass of the bins. Contours are drawn at the $\pm2-5\sigma$ levels in steps of $1\sigma$, where $\sigma$ is the rms noise in the stack. Negative contours are dashed. The colorscale of the cutouts runs from $-1.5\sigma$ to $6.5\sigma$. The number of sources in each stack is indicated in the top left corner, and was chosen to be approximately constant across the four bins. The number in the upper right corner indicates a detection (1) or upper limit (0).}
    \label{fig:mcmc_stacks}
\end{figure}

We proceed by investigating the typical level of dust obscuration in $z\sim7$ galaxies through a stacking analysis. Adopting the approach outlined in Section \ref{sec:stacking}, we randomly sample $N_\mathrm{MC} = 100$ times from the posterior stellar mass distributions of each REBELS target, and bin them into four discrete stellar mass bins. In Figure \ref{fig:fraction_in_bin} we show how the galaxies are divided among the bins across the $N_\mathrm{MC}$ samplings. The incorporation of the stellar mass uncertainties causes significant variation across the binning, with only $\sim50\%$ of galaxies being placed in the same bin $> 50\%$ of the time. We subsequently stack all the combinations of bins, and present the stacks from five randomly drawn MC runs in Figure \ref{fig:mcmc_stacks}. The highest mass bin generally appears to be the brightest, as is expected given the existence of the star formation main sequence, as well as the correlation between mass and dust obscuration (e.g., \citealt{whitaker2017}). Most of the variation is instead in the lower mass stacks, which reflects the fact that the fractional uncertainties on individual galaxies' stellar masses are larger for lower-mass sources.

\begin{figure*}
    \centering
    \includegraphics[width=0.49\textwidth]{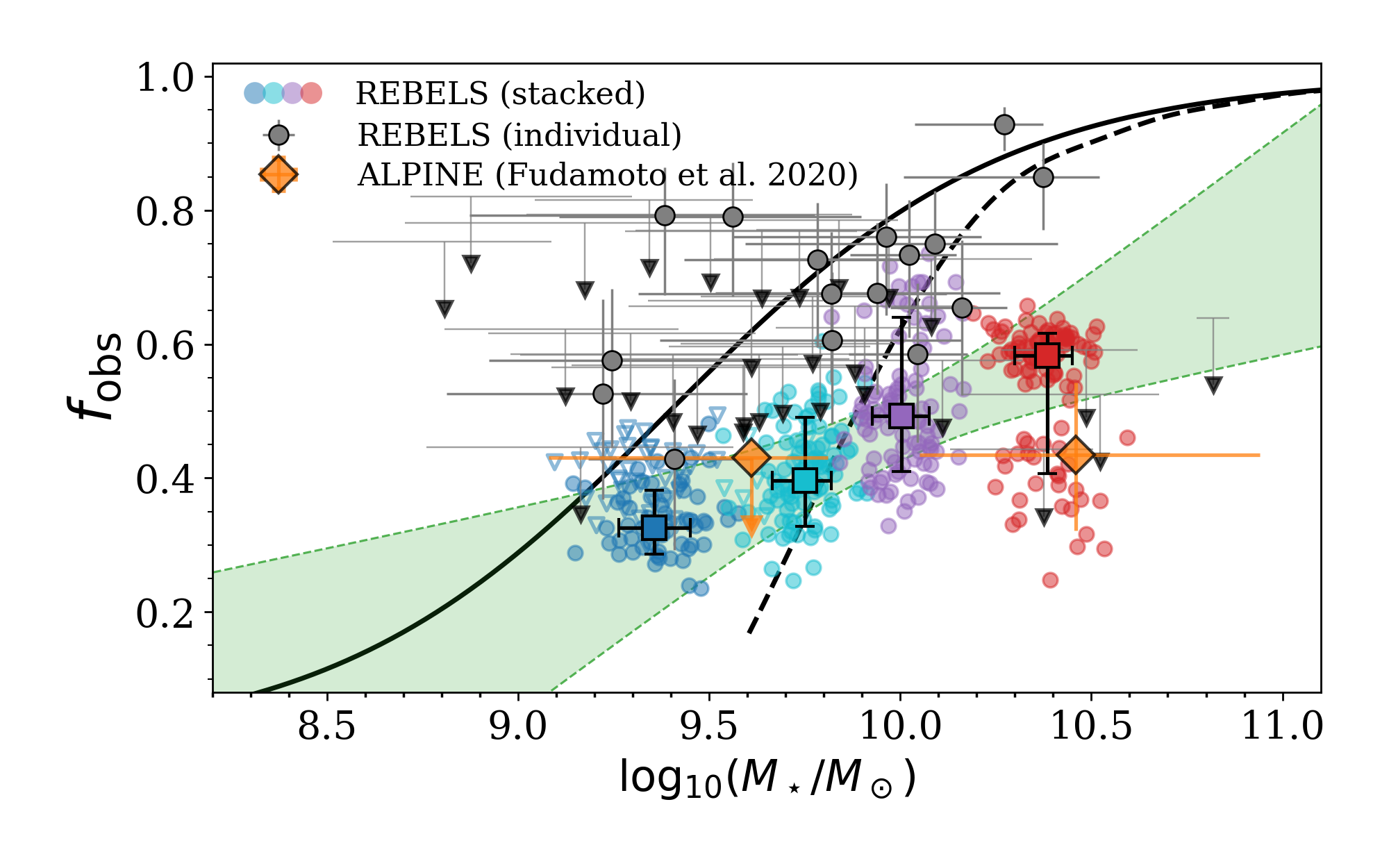}
    \includegraphics[width=0.49\textwidth]{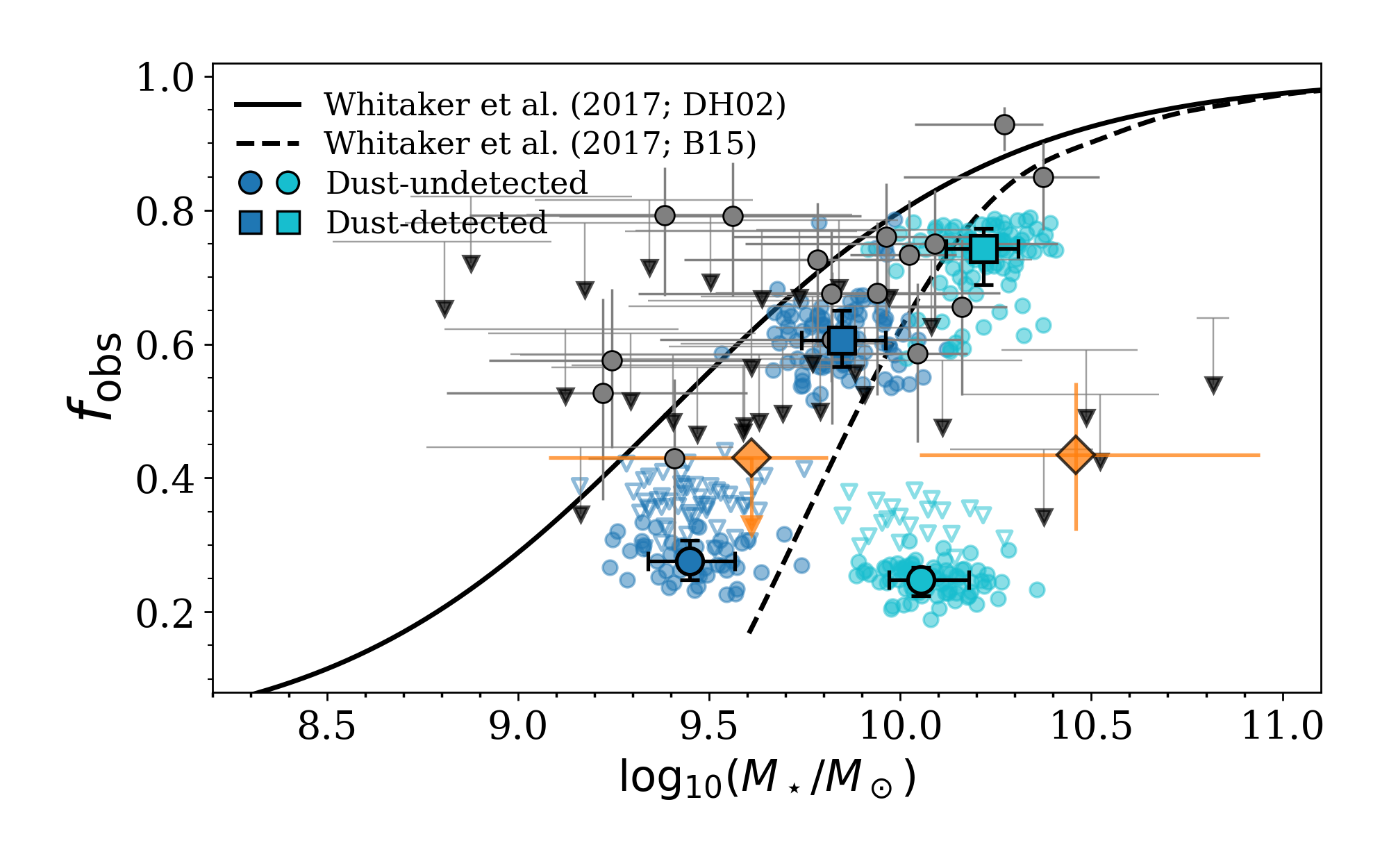}
    \caption{\textbf{Left:} The fraction of obscured star formation as a function of stellar mass. The individually detected REBELS targets as well as upper limits are shown in grey, while the Monte Carlo stacks are shown in various colors. Throughout this work, we showcase the results from the individual $N_\mathrm{MC}=100$ runs in each of four mass bins in these four colours (from low- to high-mass: blue, cyan, purple, red). Small circles (downward triangles) correspond to a single mass bin in a single MC run where the dust continuum was detected (undetected) in the stack. The larger squares correspond to the median across all $N_\mathrm{MC}$ stacks. The $1\sigma$ confidence interval of a linear fit through the individual MC samples is further shown in green. We additionally show the \citet{whitaker2017} relation determined at $z \lesssim 2.5$ (using both the \citet{dale2002} and \citet{bethermin2015} templates; solid and dashed respectively), and the $z\sim5.5$ results from \citet{fudamoto2020} as orange diamonds. While the individually detected REBELS sources agree reasonably well with the low-redshift measurements, the stacks show a clear deficit in obscured star formation. \textbf{Right:} Same as the left panel, now with separate stacks for the continuum-detected (squares) and undetected sources (circles). Both samples are split into two mass bins, and show remarkably different behaviour. The stacks of dust-detected galaxies show a typical fraction of dust obscuration that is similar to the low-redshift \citet{whitaker2017} relation, while the dust-undetected sample shows significantly less obscuration, with a typical $f_\mathrm{obs}\approx0.25$.}
    \label{fig:fobs_Mstar}
\end{figure*}

For each of the stacks we determine the fraction of obscured star formation as $f_\mathrm{obs} = \mathrm{SFR}_\mathrm{IR} / \left( \mathrm{SFR}_\mathrm{IR} + \mathrm{SFR}_\mathrm{UV} \right)$, where $\mathrm{SFR}_\mathrm{UV}$ represents the mean UV-based star formation rate of the galaxies within the corresponding bin. We show the relation between $f_\mathrm{obs}$ and $M_\star$ for the individually detected REBELS targets and the stacks -- which include both the detections and non-detections -- in the left panel of Figure \ref{fig:fobs_Mstar}. For each of the stellar mass bins, we adopt the median obscured fraction and the $16 - 84$th percentiles from the Monte Carlo stacks as the corresponding uncertainty. In the case of upper limits among the MC samples, we determine the median and uncertainties through a standard survival analysis, making use of the {\sc{lifelines}} package \citep{davidson-pilon2019}.

The obscured fraction determined for the REBELS sample ranges from $f_\mathrm{obs} = 0.33_{-0.04}^{+0.06}$ at low stellar masses ($\log_{10}\left(M_\star / M_\odot\right) \approx 9.4$) to $f_\mathrm{obs} = 0.58_{-0.18}^{+0.04}$ in the highest mass bin ($\log_{10}\left(M_\star / M_\odot\right) \approx 10.4$). We thus find that UV-selected galaxies at $z\sim7$ show an increasing obscured fraction towards higher stellar masses. Upon performing a linear fit to all $N_\mathrm{MC}$ samples and combining the posteriors, we determine a median slope and $16-84^\mathrm{th}$ percentile confidence interval of $0.27_{-0.15}^{+0.16}$ for the $f_\mathrm{obs} - M_\star$ relation. Averaged across the entire sample, the $z\sim7$ REBELS sources show a typical obscured fraction of $f_\mathrm{obs}\approx0.45$.

We first compare our results to those at $z\sim5.5$ from the ALPINE survey \citep{fudamoto2020}, and find that the obscured fractions are in agreement. Due to the typically higher continuum sensitivity of REBELS compared to ALPINE, our stacking analysis paints a clearer picture of the typical level of dust obscuration in the early Universe, including at stellar masses $M_\star \lesssim 10^{10}\,M_\odot$ where ALPINE only provides upper limits. Interestingly, REBELS galaxies -- despite residing at higher redshift -- seem to indicate a modest excess in dust obscuration at the high-mass end compared to the ALPINE sample. This is likely due to the different selection criteria of the two surveys, given that ALPINE mostly consists of galaxies on the main sequence with a priori known spectroscopic redshifts. In particular, most of these redshifts are from Lyman-$\alpha$ emission, which may result in a slight bias towards dust-poor systems as detailed in \citet{faisst2020}. REBELS, on the other hand, was selected solely based on UV brightness, irrespective of position on the main sequence, and without existing spectroscopic redshift information. As a result of these different criteria, ALPINE might preferentially miss massive, dusty galaxies.

It should be borne in mind that the ALPINE stellar masses are based on parametric SED fitting \citep{faisst2020}, while the REBELS masses use non-parametric star formation histories \citep{topping2022}. However, given the width of the ALPINE mass bins, it is unlikely that this difference in the determination of stellar masses significantly affects our comparison of the dust obscuration properties of the two samples. \\

We additionally compare our results to the relation between $f_\mathrm{obs}$ and stellar mass determined at $z\lesssim2.5$ by \citet{whitaker2017}. They find that, when compared to their high-mass counterparts, low-mass galaxies show significantly less dust obscuration. In addition, they show that the obscured fraction is independent of redshift out to $z\lesssim2.5$. However, \citet{whitaker2017} also find that the precise functional form of the relation between obscured fraction and stellar mass depends significantly upon the adopted SED templates, due to their analysis involving the extrapolation of infrared luminosities from stacked $24\,\mu$m flux densities. In particular, depending on which templates are adopted, the typical expected obscuration of galaxies with $M_\star = 10^{9.5}\,M_\odot$ may either be $\sim50\%$ (using the \citet{dale2002} templates), or negligible ($\lesssim10\%$; using the \citealt{bethermin2015} templates). However, \citet{whitaker2017} emphasize that, irrespective of which SED templates are adopted, the trend between $f_\mathrm{obs}$ and $M_\star$ remains independent of redshift. In Figure \ref{fig:fobs_Mstar}, we compare the REBELS results to the low-redshift \citet{whitaker2017} relation using both the \citet{dale2002} and \citet{bethermin2015} templates.

While the individually detected REBELS targets appear to show obscured fractions that are in reasonable agreement with the \citet{whitaker2017} relation valid at lower redshift (assuming the \citet{dale2002} templates) the stacks indicate the typical $f_\mathrm{obs}$ of $z\sim7$ galaxies is lower than observed at $z\lesssim2.5$. However, the REBELS sample shows a less significant decrease in obscured fraction towards lower stellar masses, such that the low-mass bin is in reasonable agreement with the fiducial \citet{whitaker2017} relation. When adopting the \citet{bethermin2015} templates, we find better agreement at intermediate masses, while the low-mass bin shows excess dusty star formation compared to the low-redshift relation. Compared to either of the templates, the REBELS sample shows a deficit in dust obscuration at the high-mass end. This is not due to the exclusion of starbursts from our analysis, as these sources are predominantly located at lower stellar masses. Instead, the lack of dust obscuration compared to the local \citet{whitaker2017} relation is plausibly the result of the different mechanisms of dust production at play in nearby and high-redshift galaxies. If the typical stellar populations in $z\sim7$ galaxies are young ($\lesssim 30 -40\,$Myr), dust production via AGB stars cannot contribute significantly to the overall dust mass, such that dust is likely predominantly created via Type II supernovae \citep{todini2001,bianchi2007,marassi2019,dayal2022}. In addition, the efficiency of dust build-up via grain growth in the ISM at these early cosmic epoch remains debated, although it could be important particularly in relatively massive and chemically evolved systems (e.g., \citealt{valiante2014,mancini2015,mancini2016,popping2017,graziani2020}).


However, we must also consider that REBELS, by its very nature, is a UV-selected survey. REBELS is therefore biased against selecting highly dust-obscured galaxies, and the resulting $f_\mathrm{obs}$ can hence be regarded as a formal lower limit on the true typical obscuration of $z\sim7$ galaxies. The population of highly obscured submillimeter galaxies identified at intermediate redshifts indeed forms the massive end of the galaxy population at $z\sim2-5$ (e.g., \citealt{dudzeviciute2020}), and if such a population is missed in the REBELS selection, the inferred $f_\mathrm{obs}$ at the high-mass end may therefore be underestimated. \citet{fudamoto2021} show that, even at $z\sim7$, highly dust-obscured galaxies already exist based on the serendipitous detection of two such sources in the REBELS ALMA pointings. This is further corroborated by recent work from \citet{bakx2021} and \citet{endsley2022}, who similarly find evidence of highly obscured galaxies ($f_\mathrm{obs} \sim 0.85 - 1$) in the epoch of reionization. A more complete census of dust obscuration at high redshift requires increased statistics on the prevalence of such optically faint galaxies, which are bound to arise as ALMA continues to survey the early Universe. We discuss this interesting population of optically faint/dark sources in the context of cosmic star formation in further detail in Section \ref{sec:sfrd_literature}.

\subsection{The dust-detected versus undetected population}

As expected, the left panel of Figure \ref{fig:fobs_Mstar} demonstrates that the typical obscured fraction of the dust-undetected REBELS targets is lower than that of the infrared-detected population. In order to investigate this difference between the dust-detected and undetected galaxies within our sample, we repeat our analysis by stacking these two populations separately. We adopt an identical stacking procedure as before, except we now adopt two mass bins each for both samples to ensure there is a sufficient number of sources per bin for stacking.

The $f_\mathrm{obs} - M_\star$ relation for these two populations is shown in the right panel of Figure \ref{fig:fobs_Mstar}. The dust-detected sources, which show an average $f_\mathrm{obs} \approx 0.6 - 0.7$, appear to agree quite well with the mass-dependent trend from \citet{whitaker2017}. This is somewhat remarkable, as this suggests that the brightest dust-obscured galaxies in the epoch of reionization conform to the same scaling relation as is observed at $z\lesssim2.5$, with potentially only minor evolution. However, we caveat that, while at low redshift this relation was established for a mass-complete galaxy sample, the REBELS continuum-detected galaxies are not a similarly unbiased population. In particular -- despite the detection of significant dust emission in 40\% of the sample -- REBELS may still be biased against highly dust-obscured sources given its selection at UV wavelengths. As such, we caution that a direct comparison to the \citet{whitaker2017} relation is complicated by such differences in selection.


Nevertheless, the typical obscured fraction determined here for the dust-detected REBELS sample is consistent with the modelling from \citet{ferrara2022}, who combine the UV emission and rest-frame $158\,\mu$m flux density for 10 REBELS galaxies with both dust and [CII] measurements to predict a typical obscured fraction of $f_\mathrm{obs,F22} = 0.79_{-0.19}^{+0.07}$.\footnote{Note that \citet{ferrara2022} analyze a sample of 14 REBELS galaxies, though their models are unable to provide adequate solutions for four of the targets.} In addition, theoretical modelling by \citet{dayal2022} demonstrates that the dust build-up of 11/13 of the [CII]- and continuum-detected REBELS sources can be readily explained using a combination of standard dust production mechanisms, which further supports that large dust masses, and consequently significant dust obscuration, may already be prevalent in the epoch of reionization.

In contrast to the dust detections, however, the continuum-undetected sources show significantly lower obscuration amounting to a typical $f_\mathrm{obs} \approx 0.25$. Both bins therefore fall below the fiducial $z\lesssim2.5$ \citet{whitaker2017} relation. Intriguingly, this analysis includes several galaxies with stellar masses $M_\star \gtrsim 10^{10}\,M_\odot$, none of which are individually detected in the continuum. Given that the REBELS galaxies were selected solely based on their UV luminosities, this stark difference between the dust-detected and undetected populations is interesting. \citet{inami2022} do not find significant differences in the physical properties of the continuum detections and non-detections, with the exception that the former might be characterized by redder UV-slopes $\beta_\mathrm{UV}$. A relation between the infrared excess ($\mathrm{IRX} = L_\mathrm{IR} / L_\mathrm{UV}$) and UV continuum slope has been well-established at low redshift (e.g., \citealt{meurer1999}) and provides a means of applying dust corrections when only rest-frame UV data are available (see also Bowler et al. in prep. for the $\mathrm{IRX}-\beta_\mathrm{UV}$ relation for the REBELS sample). \\

The low obscured fraction of the dust-undetected REBELS sources may have a variety of plausible origins. First of all, let us naively assume that the typical dust mass of the continuum-undetected sample is similar to that estimated for the detections by the \citet{sommovigo2022} models of $\langle M_\mathrm{dust} \rangle \approx 10^7\,M_\odot$. In this case, the continuum non-detections must have very cold dust temperatures (average $\langle T_\mathrm{dust} \rangle \lesssim 30\,$K) in order to be only weakly detected in the stacks. This, however, is unlikely, as high-redshift galaxies appear to have increasingly warmer dust (e.g., \citealt{sommovigo2022} predict $\langle T_\mathrm{dust} \rangle = 47\pm6\,$K for the REBELS targets), likely driven by compact star formation and lower overall dust masses (e.g., \citealt{schreiber2018,sommovigo2022}).

If we therefore instead assume that the REBELS continuum non-detections have a similar average dust temperature to the value determined for the detections by \citet{sommovigo2022}, they are likely to have dust masses that are lower by a factor of $\sim5\times$ compared to what is inferred for the detections. As such, it is probable that the dust-undetected population in REBELS consists of a relatively dust-poor subset of the full targeted sample of $z\sim7$ galaxies. Deeper continuum observations are required to test this hypothesis in further detail.

An alternative explanation for the differences in obscured fractions is that the UV and dust emission of the continuum-undetected REBELS sources are sufficiently spatially offset that some of the emission is diluted in the stacks. This, however, appears unlikely, as given the large beam size of REBELS ($\sim1\farcs3$) a typical offset of $\gtrsim5\,$kpc would be required. While \citet{inami2022} find that a single of the REBELS dust detections (REBELS-19) shows evidence for a large spatial offset of $\sim8\,$kpc between its UV emission and the peak of the dust, the typical separation amounts to $< 2\,$kpc ($\sim0.3'')$. As such, large ($\gtrsim5\,$kpc) spatial offsets between the UV emitting component and the dust are unlikely to be driving the lower obscured fractions.

However, more patchy obscuration on smaller spatial scales could still possibly result in a lower typical $f_\mathrm{obs}$ for the dust-undetected REBELS sample. If the distribution of the dust is sufficiently inhomogeneous, galaxies may appear more or less obscured depending on our line of sight. Higher resolution dust continuum observations are required to test this scenario in more detail.

\begin{figure*}
    \centering
    \includegraphics[width=0.8\textwidth]{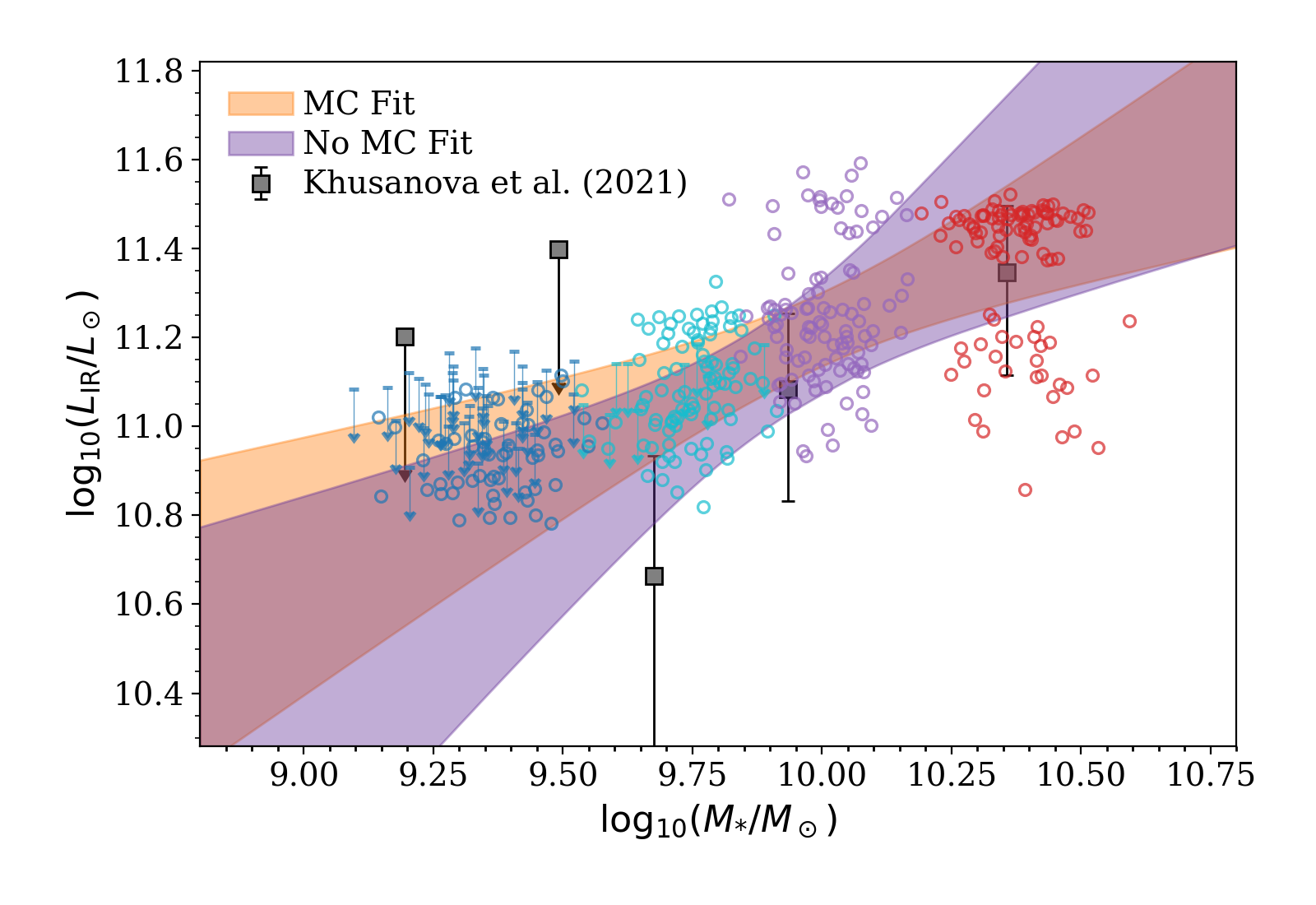}
    \caption{The relation between infrared luminosity and stellar mass for the REBELS sample. The individual colored datapoints (and upper limits) showcase the individual MC runs for each of the four bins, and are plotted at the median stellar mass of the bin. Fits to the no-MC and Monte Carlo stacks are shown in purple and orange, respectively. Including the stellar mass uncertainties through a Monte Carlo analysis results in a slightly flatter relation, though both fits are consistent within $1\sigma$, as well as with the ALPINE findings at $z\sim5.5$ (grey squares).}
    \label{fig:Lir_Mstar}
\end{figure*}

\section{The Obscured Cosmic Star Formation Rate Density}
\label{sec:results_sfrd}

Having established that dust-obscured star formation remains significant all the way into the epoch of reionization, we now set out to quantify its contribution to the cosmic star formation rate density. We remind the reader that this procedure involves constructing a relation between infrared luminosity and stellar mass, which is subsequently integrated across the known $z=7$ stellar mass function (Section \ref{sec:analysis_sfrd}). We quantify the dust-obscured SFRD following a two-step approach. First, we determine a formal ``lower limit'' on the obscured SFRD by limiting ourselves to the stellar mass range spanned by the REBELS sample. Next, we extrapolate our results to include the full range of stellar masses where dusty star formation is expected. 

Furthermore, we emphasize that our results are limited to the obscured component of cosmic star formation, and we do not explicitly use the UV SFRs of the REBELS sources in the remainder of this Section. Instead, we will use the unobscured SFRD at $z=7$ from the literature for comparison in Section \ref{sec:sfrd_literature}.

\subsection{A Lower Limit on the $z\sim7$ Obscured SFRD}
\label{sec:results_sfrd_lowlim}

We start off by determining a relation between infrared luminosity and stellar mass. We proceed with the stacked luminosities determined in the previous Section, and construct the corresponding $L_\mathrm{IR} - M_\star$ relation for all $N_\mathrm{MC}$ stacks individually (Figure \ref{fig:Lir_Mstar}). We fit the data using a standard Monte Carlo Markov Chain (MCMC) analysis, making use of the {\sc{emcee}} library \citep{foreman-mackey2013}. We adopt a flat prior on the slope of the linear fit and do not require it to be positive. We incorporate upper limits in case no stacked source was detected following the formalism outlined in \citet{sawicki2012}. We subsequently combine all MC samples across the $N_\mathrm{MC}$ stacking runs and take the median and $16 - 84$th percentiles as, respectively, the best fit and corresponding uncertainty. This results in a linear slope of the $L_\mathrm{IR} - M_\star$ relation given by $\gamma_\mathrm{MC} = 0.52_{-0.27}^{+0.29}$. We additionally determine the relation without incorporating Monte Carlo sampling across stellar masses, in which case we recover a slightly steeper slope of $\gamma_\mathrm{no-MC} = 0.74_{-0.41}^{+0.54}$. Both values are consistent with one another, as well as with the ALPINE findings at $z\sim5.5$ from \citet{khusanova2021}, who find $\gamma_\mathrm{K21} \sim 0.8 \pm 0.4$.

We integrate the $L_\mathrm{IR} - M_\star$ relations over the \citet{stefanon2021} stellar mass function at $z=7$, incorporating the covariance between their fitted parameters into our error propagation. The resulting constraints on the obscured SFRD are shown in Figure \ref{fig:sfrd} and listed in Table \ref{tab:sfrd}. We first provide a direct measurement of the SFRD by considering only the mass range spanned within the REBELS sample of $8.8 \lesssim \log_{10}\left(M_\star / M_\odot\right) \lesssim 10.8$ and adopting this as our integration limit. This, therefore, effectively constitutes a lower limit on the true dust-obscured SFRD since some fraction of cosmic star formation will also occur in galaxies outside the mass range probed by our sample. For the no-MC and MC analyses, respectively, we calculate $\log_{10}\left(\mathrm{SFRD}_\mathrm{no-MC}\right) \geq -2.74_{-0.34}^{+0.32}$ and $\log_{10}\left(\mathrm{SFRD}_\mathrm{MC}\right) \geq -2.56 \pm 0.30$ in units of $M_\odot\,\mathrm{yr}^{-1}\,\mathrm{Mpc}^{-3}$ (Table \ref{tab:sfrd}). In Figure \ref{fig:sfrd}, we adopt the lower error on these quantities (i.e., the $16^\mathrm{th}$ percentile of the posterior distribution) as the formal lower limit on the obscured cosmic SFRD.

\subsection{The Full $z\sim7$ Obscured SFRD from REBELS}
\label{sec:sfrd_full}

We next set out to determine the $z\sim7$ SFRD across the full range of stellar masses where dust-obscured star formation is likely to occur. This, therefore, requires integrating the SMF between some range $M_\star^\mathrm{lower} \leq M_\star \leq M_\star^\mathrm{upper}$ where we expect significant dust obscuration. We will show in Section \ref{sec:caveats} that the SFRD does not depend on the upper mass limit adopted for the integration, provided it is $M_\star^\mathrm{upper}\gtrsim10^{10.5}\,M_\odot$. We therefore follow \citet{khusanova2021} by adopting $\log_{10}(M_\star^\mathrm{upper} / M_\odot) = 12.4$. The value of the lower cutoff, $M_\star^\mathrm{lower}$, however, has a more significant impact on the inferred SFRD. We therefore seek to determine $M_\star^\mathrm{lower}$ in a manner that is self-consistent with our determination of the $f_\mathrm{obs} - M_\star$ relation, as outlined below. Other possible integration limits are discussed in detail in Section \ref{sec:results_sfrd_range}. 

We determine $M_\star^\mathrm{lower}$ as the mass where our $f_\mathrm{obs} - M_\star$ relation predicts that dust-obscured star formation is negligible: we first fit a linear function to the individual $N_\mathrm{MC}$ samples of the $f_\mathrm{obs} - M_\star$ relation. We adopt a flat prior on the slope, requiring it to be larger than zero. The resulting fit is shown in the left panel of Figure \ref{fig:fobs_Mstar}, and is characterized by a slope of $0.27_{-0.15}^{+0.16}$. We then use the individual MCMC samples to determine the distribution of masses $M_\star^\mathrm{lower}$ where $f_\mathrm{obs} = 0$. In the $\sim10\%$ of cases where the slope is flat, such that the inferred $\log_{10}(M_\star^\mathrm{lower}/M_\odot) < 5.0$, we adopt $\log_{10}(M_\star^\mathrm{lower}/M_\odot) = 5.0$ as our limit. The median value determined as such is $\log_{10}(M_\star^\mathrm{lower}/M_\odot) = 8.2_{-2.1}^{+0.7}$, and the full distribution of $M_\star^\mathrm{lower}$ is shown in the right panel of Figure \ref{fig:Lir_Mstar_slope}. In order to determine the total $z\sim7$ SFRD for both the MC and no-MC analyses, we randomly draw $M_\star^\mathrm{lower}$ from this distribution and integrate up to a fixed $\log_{10}(M_\star^\mathrm{upper} / M_\odot) = 12.4$. In both cases, we adopt the median and $16 - 84^\mathrm{th}$ percentile as the SFRD and its corresponding uncertainty. 

We list the $z\sim7$ obscured cosmic star formation rate densities determined via the MC and no-MC analyses across the full mass range where significant dust-obscured star formation is expected in Table \ref{tab:sfrd}, and show the results in Figure \ref{fig:sfrd}. Both approaches -- including errors on $M_\star$ or not -- provide consistent results within the uncertainties. In addition, these direct measurements of the SFRD are consistent with the lower limits derived above from considering only the range of stellar masses spanned by REBELS. In the next Section, we place our constraints on the dust-obscured SFRD at $z\sim7$ in the context of earlier studies.


\subsection{Comparison to the Literature}
\label{sec:sfrd_literature}

\begin{figure*}
    \centering
    \includegraphics[width=0.9\textwidth]{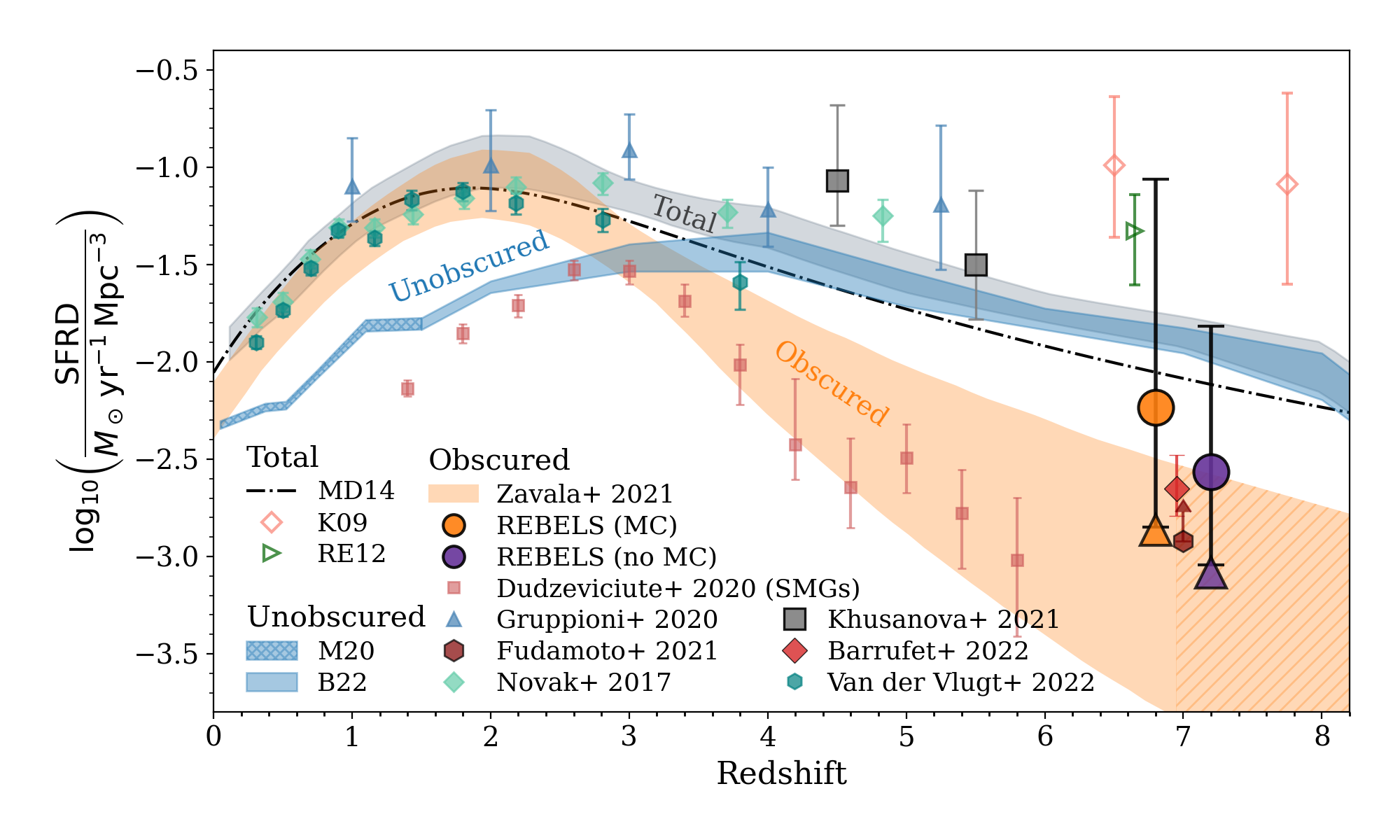}
    \caption{Constraints on the dust-obscured $z\sim7$ cosmic star formation density from the REBELS survey, using both an MC-analysis across galaxy stellar mass uncertainties (orange point), as well as a more typical approach keeping galaxy stellar masses fixed (purple). Same-color triangles indicate the lower limits obtained from considering only the REBELS stellar mass range ($8.8 \lesssim \log_{10}(M_\star/M_\odot) \lesssim 10.8$). The points are slightly offset from $z=7$ for visual clarity. We further show several SFRD measurements and compilations from the literature, divided into total, unobscured and obscured SFRD measurements. At $z \lesssim 1.5$ we show the unobscured SFRD from \citet{moutard2020} (denoted M20 in the legend), and at $z \gtrsim 2$ from \citet{bouwens2022} (B22) in blue. The dust-obscured SFRD from \citet{zavala2021} is overplotted in orange, and the combined UV + IR SFRD from these measurements is shown in grey. The hatched orange region beyond $z>7$ represents an extrapolation of the \citet{zavala2021} results to higher redshift, and is included for visual clarity. In addition, several literature results using tracers of obscured star formation are shown (far-infrared/radio; filled symbols), in addition to $z\gtrsim6$ constraints from gamma ray bursts, which are sensitive to both obscured and unobscured star formation (\citealt{kistler2009} and \citealt{robertson2012}; denoted K09 and RE12 in the legend respectively). The REBELS measurements obtained in this work suggest significant dust-obscured star formation is already in place in the epoch of reionization.}
    \label{fig:sfrd}
\end{figure*}

We compare both our lower limits and extrapolated constraints on the SFRD to various literature determinations of cosmic star formation in Figure \ref{fig:sfrd}. In particular, we compare to the models of \citet{zavala2021}, who provide the obscured cosmic star formation rate density out to $z=7$. The REBELS constraints are consistent with their obscured values, though hint towards a larger contribution of dusty star formation at this redshift. The \citet{zavala2021} models predict that most of the dust-obscured star formation beyond $z\gtrsim4$ is due to ultra-luminous infrared galaxies (ULIRGs), with an IR luminosity in excess of $L_\mathrm{IR} > 10^{12}\,L_\odot$, while only a minor fraction ($\sim20\%$) is occurring in LIRGs ($10^{11} < L_\mathrm{IR} / L_\odot < 10^{12}$). The dust-detected REBELS population, however, consists almost fully of LIRGs, with REBELS-25, attaining $L_\mathrm{IR} \approx 1.5\times10^{12}$, being the sole exception (\citealt{inami2022,sommovigo2022}; see also \citealt{hygate2022}). It therefore may be that the \citet{zavala2021} models underpredict the number density of less luminous dusty galaxies in the epoch of reionization. Through a detailed fitting procedure involving mostly observations at $z\lesssim3$, \citet{zavala2021} determine the faint-end slope of the infrared luminosity function to be relatively flat ($\alpha_\mathrm{LF}\sim -0.4$). Given their lack of constraints at higher redshift, they by necessity take this slope to be non-evolving with cosmic time. Our results suggest that $\alpha_\mathrm{LF}$ may instead evolve to steeper values, which would imply an increase in dust-obscured galaxies with LIRG-like IR luminosities in the epoch of reionization compared to their models. This is in apparent agreement with the contribution to the SFRD derived from submillimeter galaxies (SMGs) by \citet{dudzeviciute2020} across $\sim700$ ALMA-detected galaxies from the AS2UDS survey (see also \citealt{bouwens2020}). While these sources with ULIRG-like luminosities (median $L_\mathrm{IR} \approx 3\times10^{12}\,L_\odot$; \citealt{dudzeviciute2020}) contribute significantly to the total cosmic star formation rate density at $z\sim2-4$, their number density, and hence their contribution to the SFRD, rapidly drops off at $z\gtrsim4$.

In addition, the \citet{casey2018} models, upon which \citet{zavala2021} build their work, do not explicitly model the onset of dust in the early Universe. As a result, given the limited available time for dust production at $z=7$, their models may overpredict the number densities of ULIRGs with massive dust reservoirs at this cosmic epoch.

Through a fully independent approach, \citet{barrufet2022} determine the $z\sim7$ infrared luminosity function from the REBELS sample, and integrate this to obtain an estimate of the obscured SFRD. Their approach assumes the REBELS continuum detections are drawn from the known $z\sim7$ UV luminosity function, given the total area of $7\,\mathrm{deg}^2$ across which REBELS targets were initially selected \citep{bouwens2021}. The cosmic SFRD determined by \citet{barrufet2022} is consistent with the results presented here within $1\sigma$ (c.f., red diamond in Figure \ref{fig:sfrd}). This agreement between these two independent methods is encouraging, and emphasizes how valuable information on cosmic star formation can be extracted even from targeted surveys. \\

\begin{table*}
	\centering
	\caption{Constraints on the $z=7$ SFRD given a variety of assumptions and caveats, as discussed throughout Section \ref{sec:results_sfrd}. Star formation rate densities are given in units of $M_\odot\,\mathrm{yr}^{-1}\,\mathrm{Mpc}^{-3}$ and the \citet{bouwens2022} results are used to determine the obscured fraction of the SFRD ($f_\mathrm{obs}$). The lower section of the Table includes several literature $z=7$ SFRD determinations that we compare our measurements to in Section \ref{sec:sfrd_literature}.}
	\label{tab:sfrd}
	
	\def\arraystretch{1.25}
	
	\begin{tabular}{lccccl}
	    \hline\hline
	    \textbf{Description} & $\log_{10}(\mathrm{SFRD}_\mathrm{MC})$ & $\log_{10}(\mathrm{SFRD}_\mathrm{no-MC})$ & $f_\mathrm{obs,MC}$ & $f_\mathrm{obs,no-MC}$ & \textbf{Section} \\
	    \hline 
	    Fiducial & $-2.24_{-0.61}^{+1.18}$ & $-2.57_{-0.48}^{+0.75}$ & $0.31_{-0.22}^{+0.55}$ & $0.17_{-0.11}^{+0.37}$ & \ref{sec:sfrd_full} \\
	    Lower Limit & $-2.56 \pm 0.30$ & $-2.74_{-0.34}^{+0.32}$ & $0.18_{-0.08}^{+0.13}$ & $0.12_{-0.07}^{+0.10}$ & \ref{sec:results_sfrd_lowlim} \\
	    \hline\hline
	    $\gamma = 1$ & $-2.74_{-0.31}^{+0.28}$ & $-2.79_{-0.32}^{+0.30}$ & $0.12_{-0.06}^{+0.09}$ & $0.11_{-0.06}^{+0.09}$  & \ref{sec:results_sfrd_slope} \\
	    $M_\star^\mathrm{lower} = 10^{6} M_\odot$ & $-1.50_{-0.76}^{+0.93}$ & $-2.11_{-0.73}^{+0.94}$ & $0.71_{-0.41}^{+0.24}$ & $0.37_{-0.27}^{+0.47}$ & \ref{sec:results_sfrd_range} \\
	    Non-parametric SMF & $-1.82_{-0.48}^{+1.12}$ & $-2.05_{-0.36}^{+0.61}$ & $0.55_{-0.27}^{+0.39}$ & $0.41_{-0.18}^{+0.33}$ & \ref{sec:results_sfrd_mass} \\
	    Including Starbursts & $-1.54_{-0.66}^{+1.17}$ & $-1.83_{-0.55}^{+0.96}$ & $0.69_{-0.36}^{+0.28}$ & $0.54_{-0.30}^{+0.37}$ & \ref{sec:results_sfrd_starbursts} \\
	    $T_\mathrm{dust} = 35\,$K & $-2.51_{-0.56}^{+1.02}$ & $-2.78_{-0.50}^{+0.79}$ & $0.19_{-0.13}^{+0.52}$ & $0.12_{-0.08}^{+0.34}$ & \ref{sec:results_sfrd_Tdust} \\
	    $T_\mathrm{dust} = 55\,$K & $-1.51_{-0.93}^{+1.56}$ & $-1.85_{-0.89}^{+1.27}$  & $0.71_{-0.49}^{+0.28}$ & $0.53_{-0.40}^{+0.43}$  & \ref{sec:results_sfrd_Tdust} \\
	    \hline\hline
	     & \citet{bouwens2022} (UV) & \citet{fudamoto2021} (IR) & \citet{zavala2021} (IR) &  &  \\
	    $\log_{10}(\mathrm{SFRD})$ & $-1.89 \pm 0.07$ & $> - 2.92$ & $-3.19 \pm 0.66$ &  &  \\
	    \hline\hline
	\end{tabular}
\end{table*}

Our analysis suggests only a modest decline in the dust-obscured cosmic star formation rate density from $z\gtrsim3$ onward. This is in apparent agreement with recent studies of dust-obscured star formation out to $z\sim5$, such as \citet{novak2017} who make use of deep Very Large Array observations at 3\,GHz, and \citet{gruppioni2020} who utilize the serendipitous continuum-detected sources in ALPINE. Both studies find little evidence for a decline in the SFRD beyond what is commonly known as the ``peak'' of cosmic star formation, and indicate excess dust-obscured star formation compared to both the canonical \citet{madau2014} relation, and the more recent results from \citet{zavala2021}. However, recent measurements of dust-obscured star formation from \citet{vandervlugt2022}, combining both deep and wide VLA observations at 3\,GHz across COSMOS out to $z\lesssim4.5$, find a better agreement with the obscured SFRD from \citet{zavala2021}. Nevertheless, they argue UV-based surveys still underestimate dust corrections by $\sim1\,$dex, when adopting consistent integration limits for the radio and UV luminosity functions.

Indeed, the difficulty of applying dust corrections to UV-based surveys is highlighted by a galaxy population that has garnered significant attention in recent years: the so-called `optically dark' population. While the exact definition varies, these sources are characterized by their lack of a discernible counterpart at optical and near-infrared wavelengths (e.g., \citealt{wang2019,algera2020b,gruppioni2020,talia2021}), and are therefore -- by definition -- missed in UV/optical-based surveys. Based on their analysis of serendipitous sources found in the ALPINE survey, \citet{gruppioni2020} find that optically dark sources contribute about $\sim17\%$ of the overall SFRD at $z\approx5$, in general agreement with radio-selected samples \citep{talia2021,enia2022}. At $z\approx7$, \citet{fudamoto2021} recently identified two optically dark galaxies in two separate REBELS pointings, indicating that even in the epoch of reionization such highly dust-obscured sources are already in place.

As we limit our analysis to the sources directly targeted by REBELS, i.e., the UV-bright population at $z\approx7$, such a coeval optically dark population is by construction not accounted for in our obscured SFRD measurements. While detailed statistics on the nature of this population ideally require wide, blind surveys at infrared or longer wavelengths, we can provide a rough estimate on the fraction of the obscured SFRD we may be missing by comparing to the inferred SFRD from the two optically dark sources observed by \citet{fudamoto2021} at $z\approx7$ (Table \ref{tab:sfrd}; values corrected for clustering). We find that our analysis based on UV-selected sources accounts for $83_{-29}^{+16}\,\%$ of the total obscured SFRD when comparing to the estimates from \citet{fudamoto2021}, in good agreement with the $z\approx5$ results from \citet{gruppioni2020}. As such, while optically dark sources are a highly interesting population, their contribution to the obscured SFRD at $z\approx7$ might be only up to $\sim15 - 20\%$. \\

Finally, as an independent probe of the SFRD at the highest redshifts, various works have attempted to use the number counts of long-duration gamma ray bursts (GRBs) to constrain the cosmic star formation rate density out to $z\sim9$ (e.g., \citealt{kistler2009,ishida2011,robertson2012}). While subject to several systematic uncertainties -- including whether the rate of GRBs per unit star formation evolves with redshift (e.g., \citealt{robertson2012}) -- these studies can constrain the total (obscured and unobscured) star formation rate density out to high redshift. Both \citet{kistler2009} and \citet{robertson2012} suggest a significantly higher SFRD than is implied by UV-based observations alone, exceeding even our constraints from REBELS. This therefore provides independent evidence that the rate of cosmic star formation beyond $z\gtrsim5$ may be larger than was previously assumed. 

Our results thus suggest that a sizeable fraction of cosmic star formation may remain dust-obscured even at $z=7$. To quantify this, we compare to the most recent dust-uncorrected UV-based SFRD measurements at $z\sim7$ from \citet{bouwens2022}, who determine UV luminosity functions down to $M_\mathrm{UV} = -13$ aided by the gravitationally lensed galaxies observed across the Hubble Frontier Fields. We determine an obscured fraction of $f_\mathrm{obs,MC} = 0.31_{-0.22}^{+0.55}$ and $f_\mathrm{obs,no-MC} = 0.17_{-0.11}^{+0.37}$ for the MC and no-MC-based stacking analyses, respectively. These fractions are consistent with one another, and highlight how at $z\approx7$ roughly $\sim30\%$ of the total star formation rate density may be dust-obscured.
While formally consistent within the uncertainties, the obscured fraction at $z\approx7$ inferred here appears lower by $\sim2\times$ than the value determined for ALPINE at $z=5.5$ by \citet{khusanova2021}, who found $f_\mathrm{obs,K21} = 0.61_{-0.25}^{+0.20}$. This indicates that the build-up of dust rapidly follows the build-up of stars in massive high-redshift galaxies (e.g., \citealt{graziani2020,dicesare2022}).


\subsection{Caveats}
\label{sec:caveats}
We caution that our constraints on the obscured SFRD are obtained from a targeted survey, such that any selection biases will propagate into our measurement of cosmic star formation. While REBELS sources appear representative of the UV-bright population at $z\sim7$ \citep{bouwens2021}, parallel constraints from blind (sub-)millimeter surveys will be crucial to strengthen our results. In the future, such results are likely to arise from upcoming 50m class single-dish (sub-)millimeter telescopes, such as the Large Submillimeter Telescope (LST; \citealt{kawabe2016}) and Atacama Large Aperture Submillimeter Telescope (AtLAST; \citealt{klaassen2020}). In the meantime, let us discuss in detail any biases that may affect our constraints on the cosmic star formation rate density. The magnitudes of these biases, and their effect on the fraction of the SFRD inferred to be dust-obscured, are summarized in Table \ref{tab:sfrd}.

\subsubsection{The slope of the $L_\mathrm{IR} - M_\star$ relation}
\label{sec:results_sfrd_slope}

\begin{figure*}
    \centering
    \includegraphics[width=0.495\textwidth]{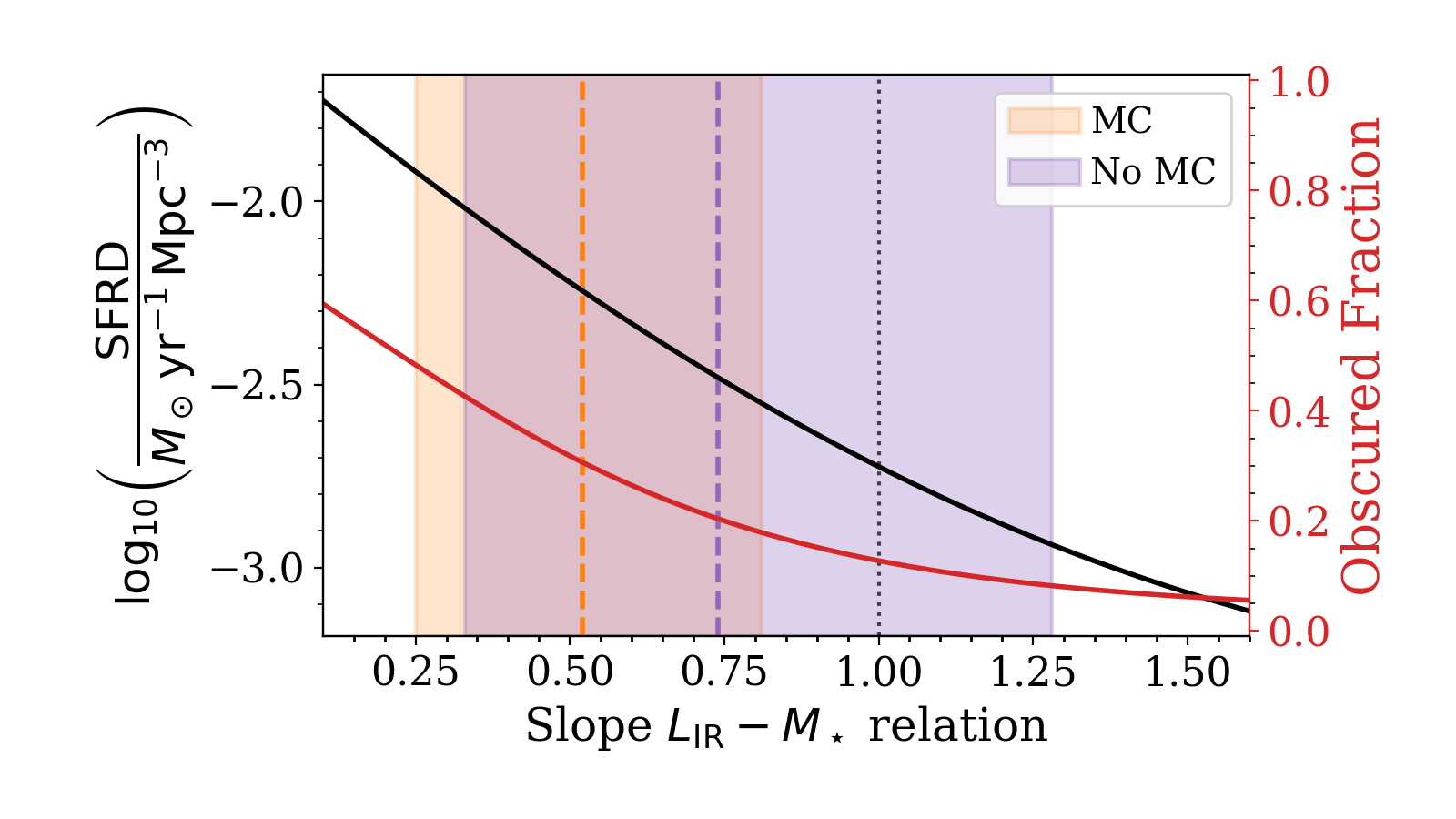}
    \includegraphics[width=0.495\textwidth]{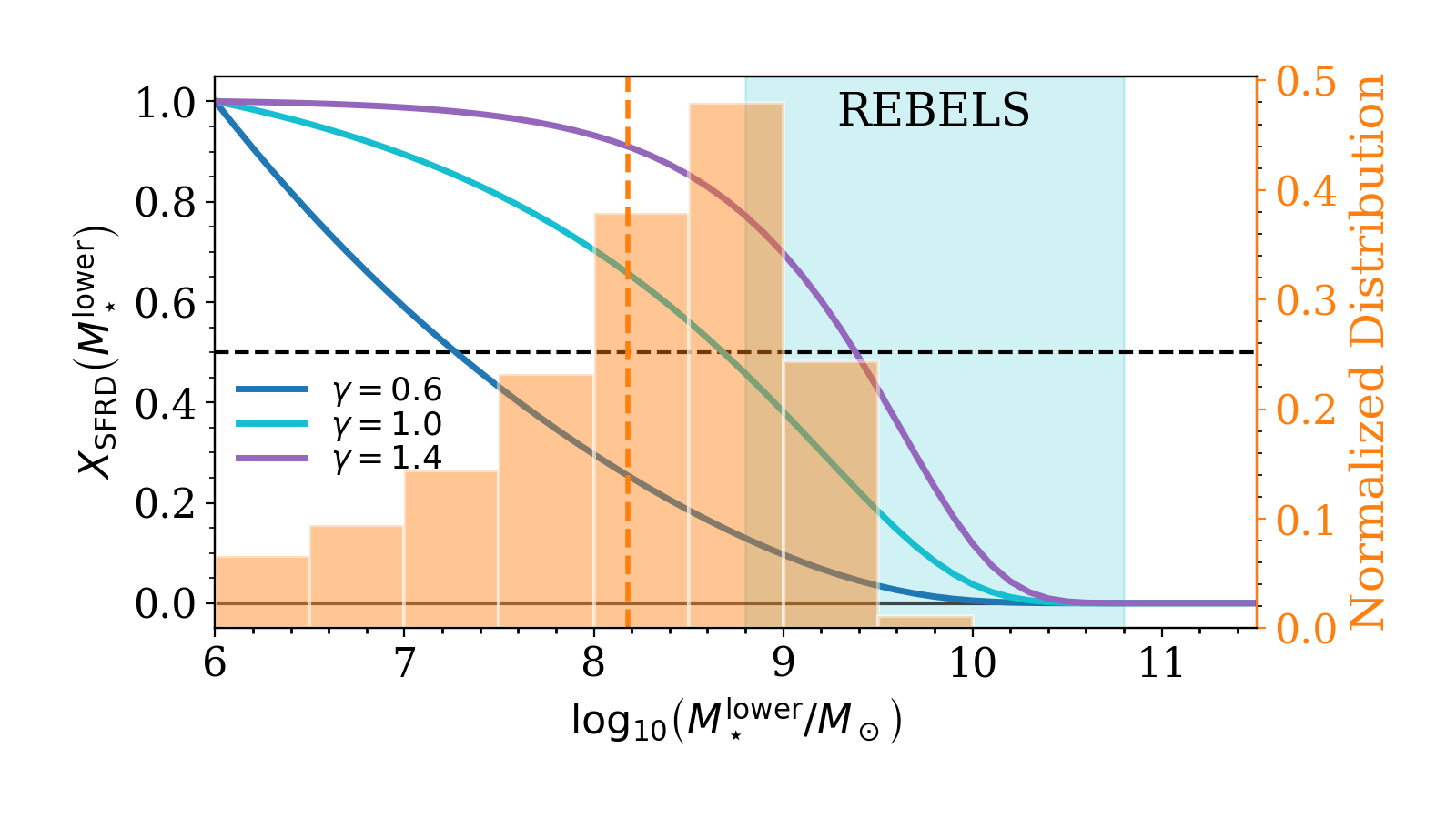}
    \caption{\textbf{Left:} The dependence of the IR cosmic star formation rate density (black line) and corresponding obscured fraction (red line; right $y$-axis) on the slope of the linear $L_\mathrm{IR} - M_\star$ relation, assuming a fixed $M_\star^\mathrm{lower} = 10^{8.2}\,M_\odot$. A flatter slope results in a larger obscured SFRD as more infrared luminosity is assigned to low-mass galaxies. As a result, the obscured fraction increases as well. The fitted slopes and $1\sigma$ confidence intervals for the no-MC and MC analyses are indicated through the dashed lines and shaded regions. A slope of $\gamma = 1$ (dotted vertical line) reduces the inferred obscured SFRD by a factor of $\sim0.5\,$dex compared to the MC-derived value. \textbf{Right}: The contribution to the SFRD as a function of $M_\star^\mathrm{lower}$, normalized such that $X_\mathrm{SFRD}(10^6\,M_\odot) = 1$. For $\gamma=1$ (cyan line), the contribution from galaxies with $10^6 \lesssim M_\star/M_\odot \lesssim 10^9$ and $M_\star > 10^9\,M_\odot$ is roughly equal, while steeper (shallower) values of $\gamma$ increase the relative importance of high-mass (low-mass) sources. The orange histogram shows the distribution of $M_\star^\mathrm{lower}$ adopted in Section \ref{sec:sfrd_full}, while the cyan shaded region indicates the mass range spanned by the REBELS sample. For clarity, we omit the $10\%$ of samples placed at the lower imposed cutoff of $\log_{10}(M_\star^\mathrm{lower}/M_\odot) = 5.0$ (see text). For an internally consistent determination of $M_\star^\mathrm{lower}$, we adopt the $f_\mathrm{obs} - M_\star$ relation to determine $\log_{10}(M_\star^\mathrm{lower}/M_\odot) = 8.2_{-2.1}^{+0.7}$ (orange dashed line).}
    \label{fig:Lir_Mstar_slope}
\end{figure*}

The slope of the $L_\mathrm{IR} - M_\star$ relation naturally propagates into the total obscured SFRD. For the MC-analysis, the slope we determine is relatively flat ($\gamma = 0.52_{-0.27}^{+0.29}$), such that considerable infrared luminosity is assigned to relatively low-mass galaxies. For the non-MC approach, the slope appears slightly steeper ($\gamma = 0.74_{-0.41}^{+0.54}$), albeit consistent with the MC-slope within the errors.

The question, then, is whether a blindly selected sample of $z\sim7$ galaxies would provide an $L_\mathrm{IR}-M_\star$ relation that is different from the one determined in this work. At present, such a sample does not exist. However, the low-redshift Universe can offer some insight, as here the star formation main sequence has been robustly measured down to low stellar masses ($\lesssim10^8\,M_\odot$ at $z\sim1$; e.g., \citealt{boogaard2018}). The combined main sequence ($\mathrm{SFR}_\mathrm{tot} - M_\star$) and mass-dependent fraction of dust obscuration ($f_\mathrm{obs} - M_\star$; e.g., \citealt{whitaker2017}) provide the quantity $\mathrm{SFR}_\mathrm{IR} = f_\mathrm{obs}(M_\star) \times \mathrm{SFR}_\mathrm{tot}(M_\star)$, which is related to the $L_\mathrm{IR}-M_\star$ relation only by a fixed scaling factor. At the high mass end ($M_\star\gtrsim10^9\,M_\odot$), the combination of these two relations points towards a slope of roughly unity between infrared luminosity and stellar mass at $z\sim1$, which is slightly steeper than what is determined from REBELS. This tentative difference could be either due to biases in our sample selection, or due to evolution of the relation towards $z\sim7$. With the present data, these possibilities can, however, not be distinguished.

Nevertheless, this exercise motivates the possibility of a steeper slope for the $L_\mathrm{IR} - M_\star$ relation. To explore the impact of the slope on the SFRD directly, we sample a range of possible slopes $\gamma \in [0.1, 1.6]$ normalized such that $L_\mathrm{IR}(M_\star = 10^{10}\,M_\odot) = 10^{11.2}\,L_\odot$, consistent with the intercept of our fitted relation (c.f., Figure \ref{fig:Lir_Mstar}). By integrating this set of $L_\mathrm{IR} - M_\star$ relations across the $z=7$ \citet{stefanon2021} stellar mass function, we illustrate how the SFRD varies depending on the assigned slope in the left panel of Figure \ref{fig:Lir_Mstar_slope}. For simplicity, we adopt a fixed $\log_{10}(M_\star^\mathrm{lower} / M_\odot) = 8.2$ in this test -- equal to the median value derived in Section \ref{sec:sfrd_full} --  while we explore the impact of $M_\star^\mathrm{lower}$ below. This analysis highlights how steepening the $L_\mathrm{IR} - M_\star$ relation from $\gamma = 0.5$ to $\gamma = 1.0$ decreases the inferred SFRD by $\sim0.5\,$dex (Table \ref{tab:sfrd}). However, this remains consistent with our current constraints on the SFRD within $1\sigma$. As such, we conclude that, while the slope of the $L_\mathrm{IR} - M_\star$ relation has some effect on the inferred SFRD, our quoted errorbars capture this source of uncertainty.

\subsubsection{The Adopted SFRD Integration Range}
\label{sec:results_sfrd_range}

In Section \ref{sec:sfrd_full} we placed constraints on the $z=7$ SFRD by integrating the SMF down to a variable cutoff mass $M_\star^\mathrm{lower}$, motivated by the finding that low-mass galaxies are generally dust poor (c.f., Figure \ref{fig:fobs_Mstar}; also \citealt{whitaker2017}). We here investigate to what extent the adopted value of $M_\star^\mathrm{lower}$ impacts the SFRD. We define the parameter $X_\mathrm{SFRD}(M_\star^\mathrm{lower})$ via 

\begin{align}
    X_\mathrm{SFRD}(M_\star^\mathrm{lower}) = \dfrac{\rho_\mathrm{IR}(M_\star^\mathrm{lower}, 12.4)}{\rho_\mathrm{IR}(6.0, 12.4)} \ ,
    \label{eq:f_sfrd}
\end{align}

\noindent where $\rho_\mathrm{IR}$ is defined as in Equation \ref{eq:rho_ir} and the numerator adopts fixed integration limits of $\log_{10}\left(M_\star/M_\odot\right) = (6.0, 12.4)$, as in \citet{khusanova2021}. As such, Equation \ref{eq:f_sfrd} measures the contribution to the obscured SFRD for galaxies with masses between $M_\star^\mathrm{lower} \leq M_\star \leq 10^{12.4}\,M_\odot$, relative to all galaxies with $10^6 \leq M_\star/M_\odot \leq 10^{12.4}$, and hence illustrates which galaxy populations host the bulk of the dust-obscured star formation.

We show $X_\mathrm{SFRD}(M_\star^\mathrm{lower})$ in the right panel of Figure \ref{fig:Lir_Mstar_slope}, and overplot the distribution of $M_\star^\mathrm{lower}$ determined empirically for the REBELS sample (Section \ref{sec:sfrd_full}). Unsurprisingly, for a shallower slope $\gamma$ the relative contribution of low-mass galaxies to the SFRD is enhanced. In particular, when $\gamma=0.6$, roughly equal to the value determined from the MC analysis, galaxies within the REBELS mass range only contribute $\sim20\%$ of the total dust-obscured SFRD, given a naive lower-mass cutoff of $M_\star^\mathrm{lower} = 10^6\,M_\odot$. As a contrasting example, if $\gamma=1.4$ is adopted -- significantly steeper than what is observed in our analysis -- around $\sim80\%$ of dust-obscured star formation is expected to take place in galaxies in the stellar mass range probed by REBELS, even for a cutoff of $M_\star^\mathrm{lower} = 10^6\,M_\odot$ (c.f., the purple line in Figure \ref{fig:Lir_Mstar_slope}).

Our analysis, however, favours a relatively shallow $\gamma\sim0.5-0.8$, which suggests that $M_\star^\mathrm{lower} \gg 10^6\,M_\odot$. Observations at lower redshift from \citet{leslie2020} indicate that at $z\sim2$ roughly half of cosmic star formation occurs in galaxies with $M_\star \gtrsim 10^{10.5}\,M_\odot$. While at higher redshift this characteristic mass decreases slightly, at $z\sim5$ \citet{leslie2020} find that half of all star formation takes place in galaxies with $M_\star \gtrsim 10^{9.7}\,M_\odot$, similar to the typical mass of galaxies in the REBELS sample. This therefore further motivates our choice of a variable $M_\star^\mathrm{lower}$. In addition, at $z\lesssim2.5$, it has been established that dust-obscured star formation is sub-dominant at low stellar masses ($f_\mathrm{obs} < 0.5$ at $M_\star < 10^{9.2}\,M_\odot$; \citealt{whitaker2017}, using the \citealt{dale2002} templates). An extrapolation of their relation to lower masses indicates that the typical fraction of dust-obscured star formation is $\lesssim10\%$ at $M_\star < 10^{8.5}\,M_\odot$. In addition, if, as in Section \ref{sec:results_fobs}, we adopt the \citet{bethermin2015} templates, we expect that considerably more massive galaxies of $M_\star \sim 10^{9.5}\,M_\odot$ already show little obscuration (typical $f_\mathrm{obs} \sim 0.1$; c.f., Figure \ref{fig:fobs_Mstar}). This is consistent with the ALMA stacking analyses of Lyman-break galaxies at intermediate and high redshifts from \citet{carvajal2020} and \citet{bouwens2020} across the Hubble Frontier Fields and Hubble Ultra Deep Field, respectively, who do not detect obscured star formation in stacks of low-mass galaxies. \citet{bouwens2020}, in particular, determine an upper limit on the obscured SFR of $0.6\,M_\odot\,\mathrm{yr}^{-1}\,(4\sigma)$ for all $1.5 < z < 10$ Lyman break galaxies with $\log_{10}\left(M_\star\right) < 9.25$, indicating the typical low-mass galaxy population does not exhibit significant dust-obscured star formation.

As such, both this work and previous studies indicate a lack of dusty star formation in low-mass galaxies at high redshift. However, for the sake of completeness, had we adopted a fixed integration range of $6.0 \leq \log_{10}\left(M_\star/M_\odot\right) \leq 12.4$, we would infer a total SFRD that is roughly $0.7\,$dex larger than the values presented in Figure \ref{fig:sfrd}. Nevertheless, this remains consistent with the constraints from our fiducial analysis, which assumes a more realistic $M_\star^\mathrm{lower}$ cutoff, within the uncertainties.


Finally, the right panel of Figure \ref{fig:Lir_Mstar_slope} shows that, irrespective of the adopted value for $\gamma$, the inferred SFRD does not depend strongly on the upper mass cutoff. Even for an assumed slope of $\gamma = 1.4$ -- steeper than inferred by our analysis -- the total amount of star formation occurring in galaxies with $M_\star \gtrsim 10^{10.5}\,M_\odot$ at $z=7$ remains negligible ($X_\mathrm{SFRD}(10^{10.5} M_\odot; \gamma = 1.4) \lesssim 1\%$). As such, we adopt an upper mass cutoff $M_\star^\mathrm{upper} = 10^{12.4}$, following \citet{khusanova2021}.

\subsubsection{Parametric vs. Non-parametric Stellar Masses}
\label{sec:results_sfrd_mass}

As already emphasized in Section \ref{sec:Mstar}, measuring accurate stellar masses is particularly challenging at high redshift due to the sparse photometric coverage of galaxy rest-frame optical emission. In this work, we adopt the non-parametric stellar masses from \citet{topping2022}, which offer greater modelling flexibility and do not require the a priori assumption of a star formation history. However, in order to determine the cosmic SFRD, we integrate across the \citet{stefanon2021} $z=7$ stellar mass function, which was constructed using parametric masses via SED modelling code {\sc{FAST}} \citep{kriek2009}.

Non-parametric star formation histories tend to result in larger inferred stellar masses compared to parametric ones (e.g., \citealt{leja2019,leja2020}). For example, comparing non-parametric and parametric masses from {\sc{Prospector}} and {\sc{FAST}}, respectively, for galaxies at $z\lesssim2.5$, \citet{leja2019} find that the former are larger by a typical $\sim0.1 - 0.3\,$dex. At higher redshift, similar results were recently obtained by \citet{whitler2022} for a sample of galaxies at $z\sim7$. In addition, \citet{topping2022} explicitly compare parametric and non-parametric mass measurements for the REBELS sample, and find that the non-parametric stellar masses from {\sc{Prospector}} exceed the parametric ones from {\sc{Beagle}} by $\sim0.43\,$dex. Upon including the nine pilot sources, and removing the four galaxies targeted in [OIII] emission (Section \ref{sec:data}), we find a similar median difference between the mass measurements of $0.45\,$dex.

\citet{whitler2022} show that the difference between parametric and non-parametric masses is a function of the inferred galaxy age, with younger galaxies being assigned comparatively higher masses when fitted with non-parametric models. While young galaxies indeed tend to be less massive, given that they are -- by definition -- in the initial stages of building up their stars, we find no evidence of a trend between $\Delta \log_{10} M_\star = \log_{10}(M_\star^\mathrm{non-par} / M_\star^\mathrm{par})$ with parametric stellar mass across the mass range spanned by the REBELS sample ($8.8 \lesssim \log_{10}(M_\star/M_\odot) \lesssim 10.8$). A linear fit between these parameters returns a slope of $-0.06 \pm 0.12$, consistent with zero within $1\sigma$. This is in qualitative agreement with the low-redshift results from \citet{leja2019}, who find only a minor trend as a function of parametric mass between $8.5 \lesssim \log_{10}(M_\star/M_\odot) \lesssim 11.5$.

To investigate the effect of integrating our non-parametric $L_\mathrm{IR} - M_\star$ relation across a parametric stellar mass function on the obscured $z=7$ SFRD, we therefore shift the \citet{stefanon2021} SMF to higher masses by a constant factor of $0.45\,$dex to mimic the assumption a non-parametric star formation history. We integrate this shifted SMF via Equation \ref{eq:rho_ir} as before to obtain the obscured cosmic star formation rate density. The resulting values exceed the SFRD inferred in Section \ref{sec:sfrd_full} by $\sim0.4 - 0.5\,$dex (Table \ref{tab:sfrd}), which is well within the uncertainties of our measurements. Given that REBELS only probes a modest (non-parametric) mass range, such a scaling of the full SMF is inherently uncertain. Further investigating the difference between parametric and non-parametric stellar mass functions is beyond the scope of this work, and as such, we opt to use stellar masses and SMFs directly from the literature. The JWST will solve some of these uncertainties by performing the much-needed robust stellar mass measurements for both our and other samples of high-redshift galaxies (e.g., GO-1626; PI Stefanon).

\subsubsection{The Effect of Excluding Starbursts}
\label{sec:results_sfrd_starbursts}

Our analysis involves determining the SFRD by first constructing a relation between infrared luminosity and stellar mass (Section \ref{sec:analysis_sfrd}), and subsequently integrating this relation across the $z=7$ stellar mass function. Given that REBELS is not an explicitly main-sequence selected galaxy sample, it contains several galaxies with SFRs in excess of the (extrapolated to $z=7$) main sequence relation (Section \ref{sec:starburst} and Figure \ref{fig:mainsequence}). Crucially, given that the continuum sensitivity of REBELS does not depend on stellar mass, REBELS is only able to detect continuum-bright galaxies at the low-mass end ($M_\star\lesssim 10^{9.5}\,M_\odot$). At the high-mass end, however, both starbursts and main-sequence galaxies are selected.

However, the existence of the main sequence at $z=7$ has not been robustly established observationally, although simulations predict it to already be in place (e.g., \citealt{ma2018}). As such, it is worth investigating quantitatively how incorporating starbursts into our analysis may affect the inferred SFRD.

Including starbursts increases our total sample by an average of $8.5 \pm 2.1$ sources, and extends the mass range of the bins down to $\log_{10}\left(M_\star / M_\odot\right) = 9.1$ (c.f., $9.4$ in our fiducial analysis). Given the addition of low-mass, infrared-bright sources, we find a shallower trend between obscured star formation and stellar mass ($f_\mathrm{obs} - M_\star$), with a slope of $0.19_{-0.10}^{+0.11}$ (c.f., $f_\mathrm{obs} = 0.27_{-0.15}^{+0.16}$). As a result, the lower mass cutoff decreases by $\sim0.8\,$dex to $\log_{10} ( M_\star^\mathrm{lower} / M_\odot) = 7.4_{-2.4}^{+0.9}$. As expected, the $L_\mathrm{IR} - M_\star$ relation also flattens upon the inclusion of starbursts, with a slope of $\gamma_\mathrm{MC} = 0.31_{-0.19}^{+0.20}$ (c.f., $\gamma_\mathrm{MC} = 0.52_{-0.27}^{+0.29}$). We note, however, that all recovered values remain consistent with their fiducial counterparts, based on excluding starbursts, within $1\sigma$. Finally, integrating this flatter $L_\mathrm{IR} - M_\star$ relation down to the newly inferred $M_\star^\mathrm{lower}$ results in an obscured SFRD that is larger than the fiducial value by $\sim0.7\,$dex (Table \ref{tab:sfrd}). This value is similar to the SFRD inferred by studies of high-redshift Gamma Ray Bursts \citep{kistler2009,robertson2012}, but remains consistent with our fiducial SFRD within $1\sigma$.

The above analysis highlights how including starbursts in the analysis increases the SFRD due to 1) a decrease of $M_\star^\mathrm{lower}$ implying an additional contribution to the infrared SFRD from low-mass sources, and 2) a flatter $L_\mathrm{IR} - M_\star$ relation increasing the typical infrared luminosity of the low-mass galaxy population. Combined, the inclusion of starbursts therefore results in a larger obscured SFRD as a result of mixing two populations in the analysis: starbursts at low masses and predominantly main-sequence galaxies at high masses.

Despite the uncertainties associated with excluding starbursts based on their offset from the extrapolated main sequence, this step is required to avoid biasing the inferred SFRD. In the near future, JWST will provide both accurate stellar masses and a clear census of the main sequence in the epoch of reionization, which will allow a more accurate investigation of what constitutes a starburst galaxy at $z\gtrsim7$.

\subsubsection{Flux Density to Infrared Luminosity Conversion}
\label{sec:results_sfrd_Tdust}

We conclude by briefly discussing how the adopted conversion between $158\,\mu$m continuum flux density and infrared luminosity affects the obscured SFRD inferred with REBELS. \citet{sommovigo2022} have analyzed the 13 REBELS sources with both a dust continuum and [CII] detection to constrain their dust temperatures (adopting a fixed dust emissivity of $\beta = 2.0$). Based on these dust temperatures, which range from $39 - 58\,$K, they provide a conversion factor from rest-frame $158\,\mu$m specific luminosity to total infrared luminosity. In their analysis of the dust continuum properties of REBELS galaxies, \citet{inami2022} adopt the median conversion factor from \citet{sommovigo2022} to compute both the infrared luminosities of the REBELS sources directly detected by ALMA, and upper limits for the rest. As such, the infrared luminosities and limits determined for the REBELS sample effectively assume a modified blackbody SED with a dust temperature of $\langle T_\mathrm{dust} \rangle = 45\pm4\,$K and $\beta=2.0$.\footnote{These values have been adapted for the non-parametric masses from \citet{topping2022}; Section \ref{sec:analysis_Lir}.} As a result, the typical uncertainty on the stacked infrared luminosities, which combines the detection significance and uncertainty on the $L_\mathrm{IR}$ conversion factor, amounts to $\sim0.15\,$dex.


To illustrate the dependence on the adopted infrared SED, we consider the effect of adopting a different average dust temperature while keeping $\beta = 2.0$ fixed. Based on the temperatures predicted by the \citet{sommovigo2022} models, there is no discernible relation between dust temperature and stellar mass for the REBELS sample. While low-mass galaxies may have hotter dust temperatures owing to their typical lower metallicities (e.g., \citealt{sommovigo2022}), this remains to be verified observationally. As such, we here assume a simple systematic under/overprediction of the typical dust temperature of the REBELS sample by $\pm10\,$K.

For an assumed temperature of $T_\mathrm{dust} = 35\,$K, the inferred obscured SFRD decreases by only $\sim0.25\,$dex (Table \ref{tab:sfrd}). This is due to two competing effects: when a lower temperature is adopted, fewer galaxies are flagged as starbursts ($6.1\pm1.8$, on average). The resulting $L_\mathrm{IR} - M_\star$ relation is therefore flatter than in the fiducial scenario, which enhances the inferred SFRD. However, the infrared luminosities of the stacks themselves decrease, given that $L_\mathrm{IR} \propto T_\mathrm{dust}^{4 + \beta}$ where $\beta=2$ is assumed. As a result the overall inferred obscured SFRD decreases slightly when a lower $T_\mathrm{dust}$ is adopted.

On the other hand, if we assume $T_\mathrm{dust} = 55\,$K, the inferred obscured SFRD increases by $\sim0.7\,$dex. This relatively large difference with respect to our fiducial determination is mainly due to the increased IR luminosities of the individual stacks. As a result, the obscured fractions increase, and hence our lower integration limit ($\log_{10}(M_\star^\mathrm{lower}/M_\odot) = 7.1_{-2.1}^{+1.0}$) decreases by roughly $1\,$dex compared to fiducial value. In addition, the infrared luminosities of the individual stacks increase when hotter dust is assumed. Nevertheless, even given $T_\mathrm{dust} = 55\,$K, the inferred obscured SFRD remains consistent with our fiducial determination within the current uncertainties. Future multi-band ALMA observations of the REBELS sources will be crucial to better constrain their dust SEDs, and to hence provide more accurate infrared luminosities for individual sources.


\section{Summary}
\label{sec:summary}

We use the dust continuum observations of a statistical sample of 45 galaxies from the ALMA REBELS Large Program (\citealt{bouwens2021}; 36 sources) and its pilot surveys (\citealt{smit2018,schouws2021,schouws2022a}; 9 sources) at $z\sim7$ with the aim of understanding the dust-obscured star formation properties of galaxies in the epoch of reionization. We divide the REBELS sample into bins of stellar mass, and adopt an image stacking analysis in the rest-frame $158\,\mu$m ALMA continuum maps in order to constrain the typical fraction of obscured star formation ($f_\mathrm{obs} = \mathrm{SFR}_\mathrm{IR} / \mathrm{SFR}_\mathrm{UV + IR}$) in UV-bright galaxies at $z\sim7$. Given the complexities of accurately measuring stellar masses at $z\sim7$, we adopt a Monte Carlo approach sampling across the $M_\star$ posterior probability distributions for the REBELS targets in order to robustly incorporate the stellar mass uncertainties into our analysis.

We find that $f_\mathrm{obs}$ increases with stellar mass, similar to what is observed at $z\lesssim2.5$ \citep{whitaker2017}. However, at fixed $M_\star$, the REBELS sources are less obscured than their lower redshift counterparts, with the obscuration ranging from $f_\mathrm{obs} = 0.3 - 0.6$ across $\log_{10}\left(M_\star / M_\odot\right) = 9.4 - 10.4$ (Figure \ref{fig:fobs_Mstar}). This apparent lack of dust obscuration may highlight differences in dust production mechanisms at high redshift, in accordance with theoretical studies (e.g., \citealt{popping2017,vijayan2019,graziani2020,dayal2022}). Interestingly, however, a more complicated picture arises when the continuum detections and non-detections are investigated separately. Through a stacking analysis, we determine the typical obscured fraction of the continuum undetected sources to be $f_\mathrm{obs} \approx 0.25$ at a median stellar mass of $\log_{10}\left(M_\star/M_\odot\right) \approx 9.8$. In contrast, the detections exhibit a typical $f_\mathrm{obs} \approx 0.70$ at $\log_{10}\left(M_\star/M_\odot\right) \approx 10.0$. As such, the continuum-detected sources show obscured fractions that are in close agreement with the $0 \lesssim z \lesssim 2.5$ results from \citet{whitaker2017}, while the non-detections contain significantly less obscured star formation. This indicates a large variety in the dust properties of $z\sim7$ galaxies, even across a sample uniformly selected based on rest-frame UV emission.

Having established the importance of dust already at $z\sim7$, we next determine the dust-obscured cosmic star formation rate density with REBELS. However, given that REBELS is a targeted survey and hence does not constitute a volume-limited sample, we follow the approach taken by \citet{khusanova2021} in order to estimate the infrared SFRD. We first establish a relation between the infrared luminosity and stellar mass of the REBELS sample (Figure \ref{fig:Lir_Mstar}), and subsequently integrate this relation across the $z=7$ stellar mass function derived by \citet{stefanon2021}. In spite of substantial uncertainties on our measurements, we find dust-obscured star formation to contribute $\sim30\%$ of the total star formation rate density at $z=7$ (Figure \ref{fig:sfrd}), highlighting the importance of dust even within the first billion years following the Big Bang.

While we note our results may be affected by selection biases in the REBELS survey, as well as any systematic uncertainties in deriving stellar masses and infrared luminosities for the sample, these sources of uncertainty are accurately captured within the quoted errorbars. In the near future, these uncertainties will be mitigated as the JWST is set to provide accurate stellar mass measurements for galaxies in the epoch of reionization. In addition, through future ALMA observations of the REBELS sample in multiple bands, the dust properties of $z\sim7$ galaxies will be mapped in ever-increasing detail, allowing for further unraveling of the importance of dust in the earliest cosmic epochs.

\section*{Acknowledgements}
We thank the anonymous reviewer for providing useful feedback that improved this manuscript.

ALMA is a partnership of ESO (representing its member states), NSF (USA) and NINS (Japan), together with NRC (Canada), NSC and ASIAA (Taiwan), and KASI (Republic of Korea), in cooperation with the Republic of Chile. The Joint ALMA Observatory is operated by ESO, AUI/NRAO and NAOJ. This paper makes use of the following ALMA data: ADS/JAO.ALMA\#2019.1.01634.L, \\
ADS/JAO.ALMA\#2017.1.01217.S, \\ADS/JAO.ALMA\#2017.1.00604.S, \\
ADS/JAO.ALMA\#2018.1.00236.S, \\ADS/JAO.ALMA\#2018.1.00085.S, \\
ADS/JAO.ALMA\#2018.A.00022.S.

We acknowledge assistance from Allegro, the European ALMA Regional Center node in the Netherlands. This work was supported by NAOJ ALMA Scientific Research Grant Code 2021-19A (HSBA and HI). We acknowledge support from the Swiss National Science Foundation through project grant 200020\_207349 (PAO, LB). The Cosmic Dawn Center (DAWN) is funded by the Danish National Research Foundation under grant No.\ 140. MS acknowledges support from the CIDEGENT/2021/059 grant, and from project PID2019-109592GB-I00/AEI/10.13039/501100011033 from the Spanish Ministerio de Ciencia e Innovaci\'{o}n -- Agencia Estatal de Investigaci\'{o}n. MA acknowledges support from FONDECYT grant 1211951, ‘ANID + PCI + INSTITUTO MAX PLANCK DE ASTRONOMIA MPG 190030’, ‘ANID + PCI + REDES 190194’, and ANID BASAL project FB210003. PD acknowledges support from the NWO grant 016.VIDI.189.162 (``ODIN") and the European Commission's and University of Groningen's CO-FUND Rosalind Franklin program. YF acknowledges support from NAOJ ALMA Scientific Research Grant number 2020-16B. IDL acknowledges support from ERC starting grant \#851622 DustOrigin.

\section*{Data Availability}
The data underlying this article will be shared on reasonable request to the corresponding author.



\bibliographystyle{mnras}
\bibliography{main} 


\bsp	
\label{lastpage}
\end{document}